\documentclass[prx,twocolumn,amsmath,noshowkeys,noshowpacs,amssymb]{revtex4-2}

\usepackage{graphicx,color}
\usepackage{amsbsy}
\usepackage{amsthm,mathrsfs,amsopn}
\usepackage{cancel}

\newtheorem{theorem}{Theorem}

\begin{document}

\title{{Persistence of chimera states and the challenge for synchronization in real-world networks}}

\author{Riccardo Muolo$^1$}

\author{Joseph D. O'Brien$^{2,3}$}

%\author{James P. Gleeson$^2$}

\author{Timoteo Carletti$^1$}

\author{Malbor Asllani$^4$}		
		\email[corresponding author: ]{masllani@fsu.edu}% Your name

\affiliation{$^1$ Department of Mathematics and naXys, Namur Institute for Complex Systems, University of Namur, rue Grafé 2, 5000 Namur, Belgium }

\affiliation{$^2$ MACSI, Department of Mathematics and Statistics, University of Limerick, Limerick V94 T9PX, Ireland}

\affiliation{$^3$ Acadia, Broadgate Quarter, One Snowden Street London, EC2A 2DQ, United Kingdom}

\affiliation{$^4$ Department of Mathematics, Florida State University, 1017 Academic Way, Tallahassee, FL 32306, United States of America}

\date{\today} % Leave empty to omit a date

\begin{abstract} 
\textbf{The emergence of order in nature manifests in different phenomena, with synchronization being one of the most representative examples. Understanding the role played by the interactions between the constituting parts of a complex system in synchronization has become a pivotal research question bridging network science and dynamical systems. Particular attention has been paid to the emergence of chimera states, where subsets of synchronized oscillations coexist with {asynchronous ones}. Such coexistence of coherence and incoherence is a perfect example where order and disorder can persist in a long-lasting regime. Although considerable progress has been made in recent years to understand such coherent and (coexisting) incoherent states, how they manifest in real-world networks remains to be addressed. Based on a symmetry-breaking mechanism, in this paper, we shed light on the {role that non-normality, a ubiquitous} structural property of real networks{, has in} the emergence of {several diverse dynamical phenomena, e.g., amplitude chimeras or oscillon patterns}. {Specifically, we demonstrate that the prevalence of source or leader nodes in networks leads to the manifestation of phase chimera states. Throughout the paper, we emphasize that non-normality poses ongoing challenges to global synchronization and is instrumental in the emergence of chimera states.}}
\end{abstract}

\maketitle

\vspace{0.8cm}

\section{Introduction}
\label{sec:intro}
\noindent

Many natural and artificial systems are made by numerous interacting entities that exhibit collective dynamics which cannot be simply inferred as the sum of the parts~\cite{Simon1991}. Such a structural pecularity has given rise to the complexity science where system{s} are modelled through networks of interacting individuals~\cite{newman2010networks}. One of the most emblematic emergent behaviors is synchronization, characterized by coherent oscillations of the basic interconnected dynamical entities, by which the system is composed \cite{pikovskij_synchronization:_2007, arenas_synchronization_2008}. Such a phenomenon is observed in many settings in nature, from the synchronous firing of neurons in the brain~\cite{Chouzouris} to the simultaneous flashing of fireflies during mating season~\cite{john_synchronous_1976}{. It has also been proven} crucial in the functioning of various human-made systems, including power grids~\cite{motter_spontaneous_2013} and communication networks~\cite{Sivrikaya_Yener}, to name a few examples.  
A prominent model to study the synchronization problem was introduced by Kuramoto~\cite{kuramoto, kuramoto_book}, based on the presence of phase variables, i.e., angles, associated to an ensemble of coupled oscillators whose dynamical behavior can be controlled by varying the coupling strength and/or the network topology. Interestingly, as the coupling strength goes beyond a given threshold or the network satisfies specific conditions in terms of links density or interactions topology, the system goes through a phase transition from an asynchronous, i.e., the angles evolutions are uncorrelated with one another, to a partial or fully synchronized regime, i.e., where the oscillators behave in unison. 

Latter, Kuramoto \& Battogtokh, observed a fascinating behavior of the model: {under very specific setting of parameters and initial conditions}, coherent and incoherent states can simultaneously coexist~\cite{chimera}. Such peculiar phenomenon, subsequently baptized \emph{chimera state} by Abrams \& Strogatz~\cite{Abrams_Strogatz}, (inspired by the mythological creature Chimera whose body was composed by parts of different animals), triggered an effervescent interest of the scientific community which continues until {the present day}. The main reason is that chimeras are one of the handful of examples where order represented by synchrony and disorder by asynchrony coexist simultaneously.
The study of chimera states has been a prolific topic of research in the past 20 years~\cite{Panaggio_2015} and many different kinds of such states have been discovered and the original idea further generalized, but two features seem to be common for all them: chimera is a long-lasting but still transient state, i.e., they fade away after a finite amount of time, and they are not robust with respect to the choice of initial conditions~\cite{zakharova_chimera_2020}. Both these aspects have inspired researchers to look for alternative ways to produce stable chimera states in a broader and less restrictive environment. {Achieving such a goal is of paramount importance in order to explain the increasingly frequent occurrence of chimera states in scenarios where coherent-incoherent patterns appear to be widespread.} The first of them is the case of brain dynamics where researchers have shown that neuronal networks manifest simultaneoulsy coherent and incoherent synchronization in the fMRI detected brain activity \cite{chimera_brain}. Another very recent result is that related to the flashing of fireflies where, contrary to common belief, partial synchrony and chimera states do exist \cite{chimera_firefly}.   

Based on these premises, in this paper we propose a theory for the emergence of coherent-incoherent patterns {grounded on non-normality, an ubiquitous structural property of} real-world networks for which the adjacency matrix (or other related operators) are (highly) non-normal, i.e., $\mathbf{AA}^T\neq\mathbf{A}^T\mathbf{A}$ \cite{Trefethen2005}. Using a systematic method recently introduced~\cite{Siebert, symm_break}, we show that robust chimera states naturally arise in empirical networks. We emphasize that such patterns are reminiscent of spectral features of real networks, directly inherited by the strong non-normality known to characterize such networks~\cite{asllani2018structure, OBrien_NN}. Significantly, non-normality is pervasive in both natural and human-made networks spanning from the microscopic (including neuronal \cite{ref68, ref69, ref70, Kaiser2006, ref54, ref66, ref67, ref68, ref28, ref43}, genetic \cite{ref55, Gerstein2012, ref59,  ref62, Sanz2011}, metabolic \cite{ref52}, protein-protein interactions \cite{Ewing2007}, etc.,) to the macroscopic world (offline and online social networks \cite{ref28, ref35, ref27, ref72, ref18, ref1, ref25, ref8, konect, Opsahl2009}, food webs \cite{Thompson2003, ref5, ref9, ref2, ref13, ref22, Bascompte2005, Ulanowicz1999, Christian1999, Goldwasser1993, Huxham1996, ref12, Dunne2008, ref21, ref36, ref73, ref74, Yodzis1998, ref32}, animal interactions \cite{ref44, ref45, ref46, ref47, ref48, ref49, ref50, ref51}, informational \cite{ref28, ref17, ref16, ref14, ref26, Milo2004}, economical \cite{ref42, ref31, ref30, ref60, ref61}, etc). {In recent years, the scientific community has recognized the importance of non-normal networks, and has highlighted their impact in the dynamics of complex systems ~\cite{Asllani2018PRE, asllani2018structure, OBrien_NN, Muolo2019, NN_stoch, baggio2020efficient, Johnson2020digraph, Muolo2021, Duan_Motter}.} It is well known that non-normality strongly influences the dynamical behavior in the linear regime. For instance, the non-orthogonality of the eigenvectors of a stable non-normal linear system drives the orbits far from the starting condition by following a transient growth which precedes the asymptotic exponential decay due to the stability assumption. In particular, it has been shown that the basin of attraction of homogeneous stationary and periodic oscillatory states drastically reduces its size, hence the transient growth and the stabilizing action of the nonlinearities provide an explanation for the emergence of new stable equilibria \cite{Asllani2018PRE, Muolo2019, Muolo2021}. Such peculiar dynamical behavior advises handling linear stability methods such as the Master Stability Function \cite{MSF} with particular care \cite{Muolo2021}. 

\begin{figure*}[t]
	\centering
	\includegraphics[width = \linewidth]{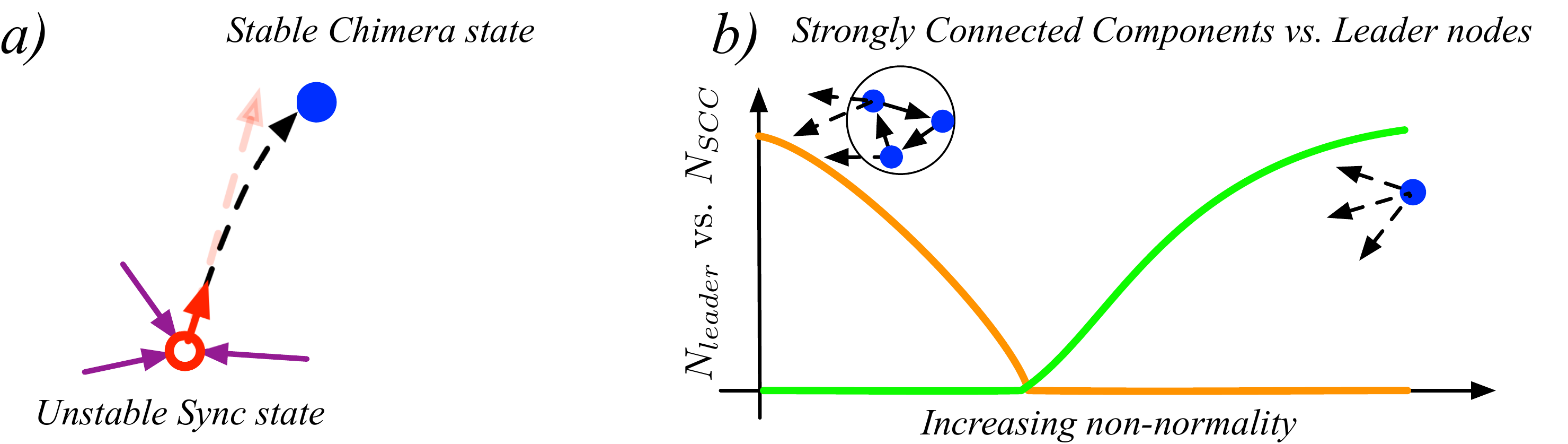}
	\caption{\textbf{Schematic representation of chimera pattern formation and leader nodes.} \textbf{a)} The process follows a symmetry-breaking mechanism where the perturbation leave the unstable synchronized manifold (red empty circle) following the direction of the critical eigenvector (red arrow {solid and shaded}) and reaches the stable chimera state (blue filled circle) through a (quasi)linear orbit (black dashed line). \textbf{b)} Once the non-normality reaches a given threshold the terminal Strongly Connected Components (SCC) (orange curve) disappear to make space to the leader nodes (green curve).}
	\label{Fig0}
\end{figure*}

The mathematical approach we use in this paper to show the omnipresence of coexisting coherent-incoherent states is rooted in the fact that the strong non-normality of complex networks is immediately related to a strong directedness of their structure \cite{asllani2018structure}, which translates into an almost triangular shape of the adjacency matrix after suitably relabeling the nodes, i.e., by performing an invariant permutation of its rows and columns \cite{asllani2018structure, OBrien_NN}. From here, it can be readily demonstrated that such an (almost) triangular matrix will also possess an (almost) triangular matrix of eigenvectors (see Methods for a mathematical derivation). Based on a symmetry-breaking mechanism (schematically represented in Fig.~\ref{Fig0} $a)$), we first show that sufficiently small perturbations around a fully synchronized unstable state, evolve from it following in the linear regime the unstable eigenvectors, namely the ones corresponding to the unstable modes. Before we enter into the details of the method, we want to emphasize that our definition of chimera state will be based on the presence of at least a subset of nodes of ordered behavior and at least a subset with a disordered one. In this paper by \emph{ordered}, we mean that some of the network's nodes share simultaneously the same values of a specific observable, either the amplitude or the phase, defined accordingly to the case under consideration. %To exemplify, if a subset of the nodes have the same uniform steady state, namely the same amplitude and no phase variable associated, with the remaining nodes having randomly distributed values while still being stationary, i.e., different amplitudes and no phases, then we talk about a chimera state. 
In the following we will consider only cases where the network is split into at least one cluster of nodes with an oscillatory state and the remaining ones with a different dynamics. Now coming back to our problem, due to the nearly triangular structure of the matrix operators, the entries of its eigenvectors contain many zeros and generally randomly scattered non-zero entries. Hence the perturbations corresponding to the zero entries, will remain near the original limit cycle while the others grow substantially with considerably different amplitudes. Once the amplitudes of these nodes reach a certain threshold, the nonlinear terms enter into action by ``freezing'' such linear growth and consequently giving life to the amplitude chimera state \cite{amp_chimera}, characterized by the presence of in-phase oscillations but with significantly different amplitudes. The merits of the approach is that the final patterns, {alongside being predictable,} are also stable, following known results in pattern formation theory and at the same time require only setting the parameters of the model but otherwise are independent of the initial conditions \cite{symm_break}.  

%\onecolumngrid

%\twocolumngrid

Non-normality in real complex systems manifests itself in other forms characterized by particular local structural features. In~\cite{OBrien_NN}, authors have shown that once the network non-normality reaches a given threshold, all the terminal Strongly Connected Component (SCC) within the system are replaced by \emph{leader} nodes, namely nodes with only incoming (sink) or outgoing (source) edges (see Fig. \ref{Fig0} $b)$). The coexistence of both strongly synchronized clusters (SSCs) and leader nodes has been found to be highly improbable in all 124 networks studied in the dataset, with only one exception \cite{OBrien_NN}. Therefore, we consider the presence of leader nodes as a distinctive characteristic indicating strong non-normality. Based on this concept, we show that in a strongly stable regime, source (leader) nodes will retain any initial phase perturbation assigned to them as being directly uncoupled from the rest of the network. The remaining nodes instead may potentially absorb not only all the perturbations they had initially, but also those coming from the source nodes. Consequently, leader nodes will act as a set of phase-disordered oscillators, while the remaining nodes in the network may eventually form a synchronized cluster. In this scenario, all nodes will maintain a constant amplitude without undergoing any changes.

The coexistence of partially organized patterns similar to the chimera scenario has been previously observed also on continuous support. This is the case of oscillon patterns which occur in granular media and are basically localized oscillatory {particles}, non necessarily synchronized states, surrounded by a neighborhood of uniform stationary ones \cite{umbanhowar_localized_1996, Vanag_Epstein}. They have also been observed in networked systems \cite{oscillon_net}. In this paper, we show that such states arise naturally in real networks, however, in constrast with the previous scenario where node dynamics are oscillatory, we hereby assume the nodes' variables initially sit on a stationary equilibrium. Through the same mechanism described above, following an oscillatory instability the nodes of the network will begin to oscillate and then stabilize because of the nonlinearities. {The} contribution of the non-normal network is given by the {eigenvectors of the coupling matrix}  which will localize such oscillations in a subset of nodes, corresponding to the non-zero entries, resulting in the emergence of an oscillon state. 

Before we move to the mathematical treatment of our study, we find it necessary to emphasize that while we predominantly employ a symmetry-breaking method to address the emergence of chimera states in this paper, the results extend qualitatevely similarly for parameters beyond the technical validity of the method. In the Supplementary Material (SM), we show that even for larger perturbations, distinct chimera patterns resembling the shape of the unstable eigevectors can still occur. 

%The paper is organized as follows: in Section \ref{sec:I} we will introduce the symmetry-breaking method and prepare the ground for Section~\ref{sec:II} where we show how amplitude chimera states systematically form in real networks. We will briefly extend our result to the case of oscillons in Section~\ref{sec:III}. In Section~\ref{sec:IV}, we show that leader nodes can potentially produce phase chimera patterns before concluding and summerizing our results. 

%%%%%%%%%%%%%%%%%%%%%%%%%%%%%%%%%%%%%%%%%%%%%%%%%%%%%%%%%%%%%%%%%%%%%%%%%%%%%%%%%%%%%%%%%%%%%%%%
%%%%%%%%%%%%%%%%%%%%%%%%%%%%%%%%%%%%%%%%%%%%%%%%%%%%%%%%%%%%%%%%%%%%%%%%%%%%%%%%%%%%%%%%%%%%%%%%

\section{Symmetry-breaking method}
\label{sec:I}
\noindent

We start by considering a coupled autonomous dynamical systems composed by $N$ variables $\mathbf{x}_i=\left(x_i^{(1)}, x_i^{(2)},\dots,x_i^{(N)}\right)$, hereby referred as observables or species, associated to the $i$-th node, for $i=1,\dots,\Omega$ and evolving because of the coupling given by:
\begin{eqnarray}
&&\dfrac{d\, x_i^{(\gamma)}}{dt}= f_{\gamma}(\mathbf{x}_i) + D_{\gamma}\sum_{j=1}^\Omega A_{ij} \left(x_j^{(\gamma)} - x_i^{(\gamma)}\right),\\\nonumber\\ &&\forall i=1,\dots,\Omega\,\,\, \mathrm{and}\,\,\, \forall \gamma=1,\dots, N\nonumber \label{eq:gen}
\end{eqnarray}
where $f_{\gamma}(\cdot)$ stands for the nonlinear term of interaction of the $N$ variables associated to the $\gamma-$th equation, $A_{ij}$ represents the $(i,j)$ entry of the adjacency matrix encoding the coupling topology, i.e., $A_{ij}=1$ if there is an edge from $j$ to $i$ and zero otherwise, and $D_{\gamma}$ describes the coupling strength. Throughout this paper we will consider identical node dynamics, e.g., identical oscillators for the whole network, although further progress to the case of non-identical oscillators is, in principle, possible \cite{Sun_2009, Zhang_2018}. {The linear coupling between nodes resembles, for instance, the gap junction in neuronal systems where the observables tend to minimizes the differences between each other \cite{varshney_structural_2011, sorrentino_synchronization_2012}.} In a compact form we can write such coupling through the graph Laplacian as $\sum_j L_{ij}x_j^{(\gamma)} = \sum_j A_{ij}\left(x_j^{(\gamma)}-x_i^{(\gamma)}\right)$ with entries $L_{ij}=A_{ij}-k_i^{in}\delta_{ij}$ where $k_i^{in}$ is the in-degree, i.e., the number of incoming links of node $i$.

The symmetry-breaking method consists on performing a linear stability analysis in order to study either the amplification or damping of a small perturbation starting from a uniform equilibrium state. The initial state can either be a limit cycle, to which all nodes are synchronized, or a homogeneous fixed point. For generality, we will consider here the case of a homogeneous periodic solution for all the nodes $x_i^{(\gamma)}(t)=\hat{x}^{(\gamma)}(t)$ and then linearize the system around it. However, it is always possible to constrain such assumptions to the case of a stationary uniform steady state.
If we denote the perturbations by $\delta x_i^{(\gamma)}(t)$ then the linearized equations read as:
\begin{eqnarray}
&&\dfrac{d\,\delta x_i^{(\gamma)}}{dt}=\sum_{\eta=1}^N \frac{\partial f_{\gamma}}{\partial x^{(\eta)}}\delta x_i^{(\eta)} + D_{\gamma}\sum_{j=1}^{\Omega}L_{ij} \delta x_j^{(\gamma)},\\\nonumber\\ &&\forall i=1,\dots,\Omega\,\,\, \mathrm{and}\,\,\, \forall \gamma=1,\dots,N\nonumber
\label{eq:linearized}
\end{eqnarray}
where we have now made use of the Laplacian notation and the partial derivatives $\partial f_{\gamma}/\partial x^{(\eta)}$ are calculated on the homogeneous periodic solutions $\hat{x}^{(\gamma)}(t)$. %\teoadd{We will refer to the matrix with entries the sum of the partial derivates and the corresponding Laplacian entries as the extended Jacobian matrix $\boldsymbol{\mathcal{J}}(t)$.} 
{To make analytical progress we will decouple the above equations and for this we need to consider an expansion of the perturbations along the eigenvectors of the Laplacian matrix, i.e., $\delta x_i^{(\gamma)}=\sum_{\alpha=1}^{\Omega} \xi_{\alpha}^{(\gamma)}(t)\Phi_i^{(\alpha)}$. The latter operation is only possible when a basis of Laplacian eigenvectors exists \footnote{However, the analysis can be extended to cases where the Laplacian matrix is not diagonalizable by utilizing Jordan blocks \cite{nondiagonal}.}; let us observe that this is always the case for a normal matrix where the eigenvectors are orthogonal, but might not be necessarily true for the non-normal ones. We come back to this problem again in the SM, but for hereby we assume that although non-orthogonal, the eigenvectors are still linearly independent. This yields to the variational equations:
\begin{eqnarray}
&&\dfrac{d\,\xi_\alpha^{(\gamma)}}{dt}=\sum_{\eta=1}^N \frac{\partial f_{\gamma}}{\partial x^{(\eta)}} \xi_\alpha^{(\eta)} + D_{\gamma} \Lambda^{(\alpha)} \xi_\alpha^{(\gamma)},\\\nonumber\\ &&\forall \alpha=1,\dots,\Omega\,\,\, \mathrm{and}\,\,\, \forall \gamma=1,\dots,N\nonumber
\label{eq:MSF}
\end{eqnarray}
where now the dynamics of the nodes are uncoupled.}

{In general, unless we perturb a fixed point, for each value of $\alpha$ we will be dealing with a time-dependent Jacobian making this way the system non-autonomous. Hence, to study the stability of each linear system~(3), we need to resort to the Maximum Lyapunov Exponent (MLE) which consists of finding the value of $\lambda_{\alpha}$ for which the evolution of the linearized system $\xi_{\alpha}^{(\gamma)}(t)$ fits an exponential trajectory $c_{\alpha}^{(\gamma)}e^{\lambda_{\alpha}t}$ where $c_{\alpha}^{(\gamma)}$ is the initial value for each of the species involved.}

{For the case of a fixed point instead, we look for an expansion of} the form $\delta x_i^{(\gamma)}=\sum_{\alpha=1}^{\Omega} c_{\alpha}^{(\gamma)}e^{\lambda_{\alpha}t}\Phi_i^{(\alpha)}$ where $\lambda_{\alpha}$ is the growth (decay) rate. Using the above expansion and collecting the terms corresponding to each eigenvector we obtain the following eigenvalue problem:
\begin{eqnarray}
&&\!\begin{pmatrix}
\dfrac{\partial f_1}{\partial x^1} + D_1\Lambda^{(\alpha)} - \lambda_{\alpha} & \dots  & \dfrac{\partial f_{1}}{\partial x^{N}}\\ \vdots &  \ddots &  \vdots\\ \dfrac{\partial f_{N}}{\partial x^1} & \dots  & \dfrac{\partial f_N}{\partial x^{N}}+ D_N\Lambda^{(\alpha)} - \lambda_{\alpha}
\end{pmatrix}\!\!\! \begin{pmatrix}
c_{\alpha}^{(1)}\\\vdots\\c_{\alpha}^{(N)}
\end{pmatrix}\!\!\nonumber\\\nonumber \\ && \,\,\,\,\,=0, \,\,\,\,\,\,\,\,\,\,\forall \alpha=1,\dots,\Omega
\label{eq:master}
\end{eqnarray}
where $\Lambda^{(\alpha)}$ represents the Laplacian $\alpha-$th eigenvalue, i.e., $\sum_{\alpha} L_{ij} \Phi_j^{(\alpha)}=\Lambda^{(\alpha)}\Phi_i^{(\alpha)}$. {Notice that, as before, problem~(2) transformes into solving  $\Omega$ independent linear systems~\eqref{eq:master}.}%, i.e., the number of nodes in the network. 

The linear stability approach followed so far is known in network science as Master Stability Function \cite{MSF}, and as the name implies, it has primarily been utilized to study the conditions under which the synchronized manifold remains stable. However, recently it has been used to achieve the opposite goal, i.e., to establish the conditions for which such stability is broken in such a way to give rise to new interesting states such as cluster synchronization, chimera states or modular Turing patterns \cite{Siebert, symm_break}. In practical terms, if a single eigenvalue (respectively Lyapunov exponent) takes a {(small)} positive value, let's say $\lambda_M>0$, then the perturbation will initially exponentially grow with an amplitude weighted by the entries of the corresponding eigenvector $\Phi_i^{(M)}$. Simultaneously, the nonlinear terms also grow, leading to the saturation of the pattern that has already formed in the linear regime. Additionally, multiple unstable modes will concurrently contribute and compete with each other in shaping the final patterns. {This approach is grounded on established results of weakly nonlinear analysis of pattern formation when the perturbed states are either fixed points or limit cycles and is based on a multiple-scale perturbative analysis. It can be shown that under the constraint of being in close proximity to the bifurcation threshold \cite{contemori_multiple-scale_2016} (and also the uncoupled limit cycles should be similarly close to the threshold due to being fixed points) a normal form known as the Ginzburg-Landau equation is obtained for the weakly coupled dynamical systems \cite{cross_pattern_2009, nakao_complex_2014}. It describes the amplitude evolution of the patterns and corresponds to a supercritical pitchfork bifurcation. Throughout this paper we will simple refer to such results paying careful attention to satisfying the necessary conditions without recalling the mathematical details (the interested reader can refer to the literature cited hereby) \cite{cross_pattern_2009, kuramoto_book, nakao_complex_2014, contemori_multiple-scale_2016, di_patti_ginzburg-landau_2018}.} 

Another advantage of this method is that, due to the continuity of the phase transition (i.e., supercritical bifurcation) the nonlinear heterogeneous states formed at the new equilibrium, are expected to be not only similar to the eigenfunctions of selected modes, but also stable in their own right \footnote{{Adding more rigor to this statement, for the case of a continuum medium, it can be shown and computed numerically that in the domain formed by the parameters and the wave vector, it exists a contiguous region, known as the ``stability balloon'' where the heterogeneous pattern is stable \cite{cross_pattern_2009}.}}. {This} is at odds with the ``classic'' chimeras in that they are stable rather than transitory and independent of the initial conditions, i.e., once the parameters are set selecting the unstable modes, any perturbation sufficiently small for the sake of the linearization approach yields robust chimera patterns. In the following sections, we will analyze various scenarios in which this method allows us to understand the emergence of different chimera states, driven by the presence of non-normality. 

{The symmetry-breaking method described in this section is pivotal for generating chimera patterns. However, it is restricted to a certain regime of parameters. In SM we show that same result extends also to the case where such restrictions are removed. For instance, when the perturbations are moderately small (but larger than those considered in this paper) the MSF is no longer effective in describing the evolution of the state of the system to the asymptotic case due to the tiny basin of attraction of the synchronized manifold (see \cite{Asllani2018PRE, Muolo2019, Muolo2021}). In this case, the MSF takes negative values and the instability is driven by the non-normal dynamics. Specifically, there is transient growth for a short period of time, which can potentially lead to a permanent instability in the system. Although we can no longer predict the exact form of the final state, as we shall see in the SM (e.g., Fig. SM1), amplitude chimera patterns still persist in the large time regime.}

%%%%%%%%%%%%%%%%%%%%%%%%%%%%%%%%%%%%%%%%%%%%%%%%%%%%%%%%%%%%%%%%%%%%%%%%%%%%%%%%%%%%%%%%%%%%%%%%
%%%%%%%%%%%%%%%%%%%%%%%%%%%%%%%%%%%%%%%%%%%%%%%%%%%%%%%%%%%%%%%%%%%%%%%%%%%%%%%%%%%%%%%%%%%%%%%%

\section{Emergence of amplitude chimera states}
\label{sec:II}
\noindent

\begin{figure*}
	\centering
	\includegraphics[width = \linewidth]{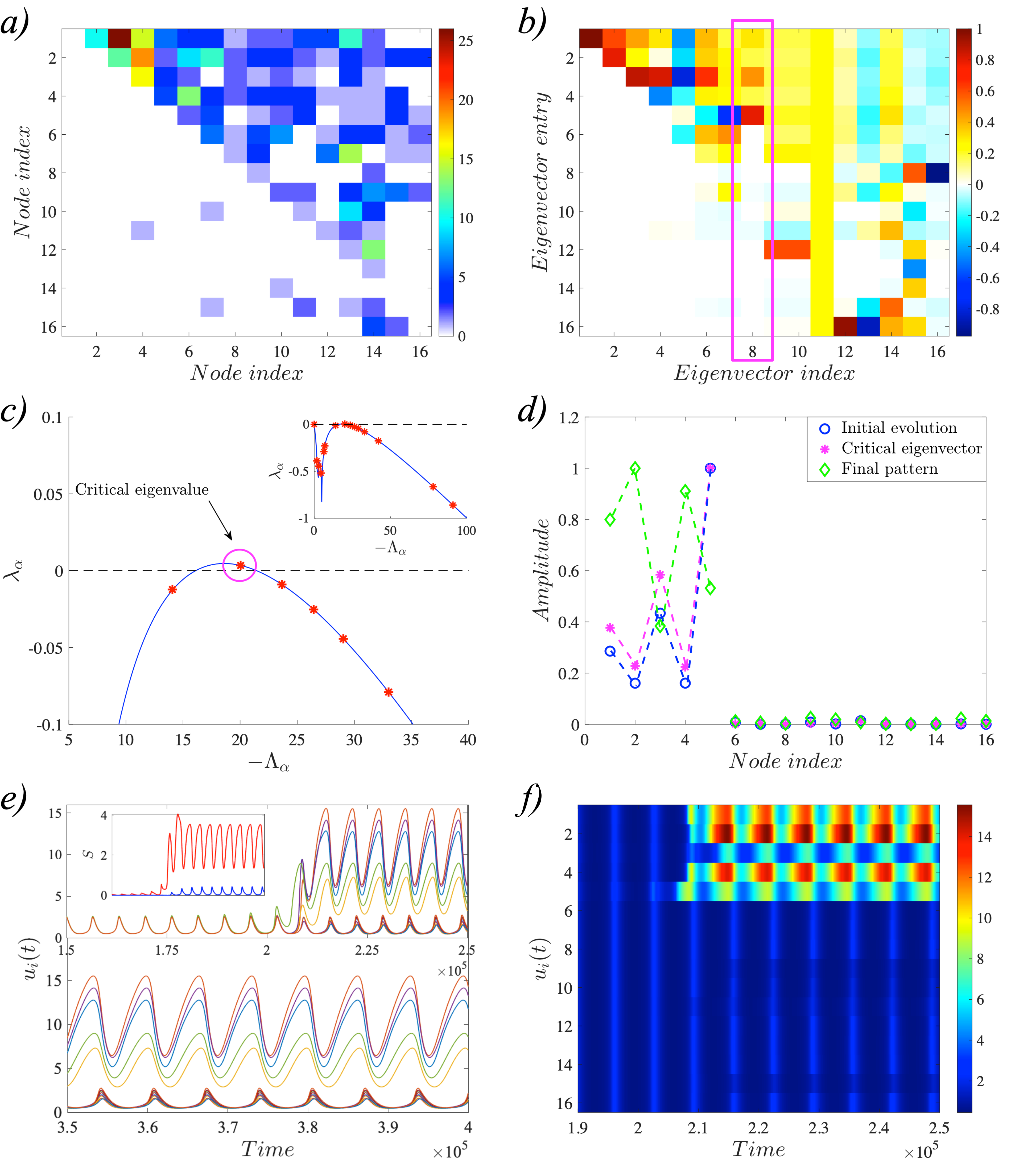}
	\caption{\textbf{Emergence of amplitude chimera state.} $\textbf{a)}$ The almost triangular adjacency matrix of the \textit{macaques competition} network \cite{ref41} ordered hierarchically. $\textbf{b)}$ The matrix where the columns are the Laplacian eigenvectors of the same network where the magenta rectangle shows the critical eigenvector. $\textbf{c)}$ The Master Stability Function (MSF) zoomed around the critical eigenvalue (the whole MSF is shown in the inset). $\textbf{d)}$ The comparison of the critical eigenvector (magenta stars) vs. the normalized amplitudes of the initial evolution (blue circles) and the final pattern (green diamonds). {With the aid of the dashed lines, it is possible to notice} that the shape of the final pattern is flipped (due to the choice of snapshot time) compared to the critical eigenvector, but otherwise is similar to it. $\textbf{e)}$ The time series for each oscillator (zoomed, lower part and the complete evolution, upper part). In the inset of the upper part the standard deviation for the first $4$ oscillators (red curve) and the last $12$ ones (blue curve) is shown. $\textbf{f)}$ The colormap representation of the oscillators dynamics evolution. The parameters for the Brusselator are $b=2.5$, $c=1$, $D_u=0.0168$, and $D_v=0.2112$ and the colorbars quantify either the magnitudes of the matrices entries or the oscillators amplitudes.}
	\label{Fig1}
\end{figure*}

Building upon the foundations established in the previous section, we will now investigate the formation of amplitude chimera patterns in a specific network selected from a set of empirical ones \cite{OBrien_NN}. An amplitude chimera is a hybrid state characterized by a mixture of behaviors. In this state, all the oscillators within the network exhibit synchronization in terms of their phases (and consequently frequencies). However, this order is disrupted when considering the amplitude variable, as the network divides into subsets of nodes that share either the same or different magnitudes of oscillations. Such states have been first discovered by Zakharova et al., for the case of coupled Stuart-Landau oscillators on a symmetric network \cite{amp_chimera}. In the next section we will present numerical results obtained for the Brusselator model with a {(weakly)} diffusive coupling. Such a model is characterized by the following local (node) dynamics: 
\begin{equation}
\begin{cases}
f(x_i,y_i)=1-(b+1)x_i+cx_i^2y_i\\
g(x_i,y_i)=bx_i-cx_i^2y_i
\end{cases}
\label{eq:Bruss}
\end{equation} 
where $b$ and $c$ are positive parameters. For simplicity of notation we have renamed the nonlinear terms as $f(\cdot, \cdot)$ and $g(\cdot, \cdot)$ and the two involving variables as $x$ and $y$, respectively. The Brusselator model can exhibit either a fixed point or a limit cycle. Assuming that we are considering the latter regime, we set the model parameters (see Fig. \ref{Fig1}) in a way that selects a single unstable mode and then trigger the system with random perturbations that are transverse to the limit cycle homogeneous equilibrium, i.e., the eigenvector $(1,\dots,1)^\top$. With this in mind, we now focus on analyzing the structure and dynamics in the \emph{macaque competition} network \cite{ref41} whose adjacency matrix, as shown in Fig. \ref{Fig1} panel $a)$, is very close to upper triangular. It is important to note that as a preliminary step we perform node relabeling on this network by appropriately permutting the row and the columns, in order for the network to follow a hierarchical structure, using a method developed in Refs. \cite{OBrien_NN, asllani2018structure}. It consists in assigning the first labels to the sink (resp. source) nodes and then assigning successive labels to the nodes immediately connected to them through incoming links. This process repeats for the remaining nodes following the hierarchy of the Directed Acyclic Graph (DAG) structure. The edges of the network are thus organized in such a way to fill the upper triangle of the adjacency matrix. Once the pool of nodes is exhausted, we take into account the remaining links which disrupt the perfect DAG structure and these are placed on the lower triangular part. Such distinct features of the network structure will significantly impact almost entirely on the Laplacian matrix whose difference with the corresponding adjacency matrix consists only on the diagonal. The {key ingredient such a particular feature of the structure adds to the symmetry breaking method,} is based on a simple mathematical fact: the eigenvectors of a triangular matrix, once permuted in the proper way, will form the columns of a matrix which is also triangular (see Methods). Consequently the expectation is that a non-normal network whose adjacency or other matrix-related operators (e.g., Laplacian) are almost triangular will also have an almost triangular eigenvectors matrix. Such an assertion is validated numerically in the Supplementary Material considering many different examples of real-world networks. This is also the case of the network under consideration as can be clearly noticed in panel $b)$ of Fig. \ref{Fig1}. Observe in particular, the uniform entries of the eleventh eigenvector indicate that it corresponds to the zero eigenvalue of the Laplacian. Once we have isolated a single unstable mode in the Master Stability Function (panel $c)$) corresponding to the eighth eigenvector (indicated by the magenta box), the evolution of the pattern will initially follow closely that of the critical eigenvector, as depicted in panel $d)$. The shape of the final pattern will be determined by the combined contribution of linear and nonlinear terms {and although we have deliberately chosen the parameters that are close to the requirements of the method (but not excessively close to challenge the limits of our approach!), the nonlinear pattern exhibits a remarkable resemblance to the unstable eigenvector. The main difference is that the symbols in the current snapshot are flipped compared to the eigenvector entries due to the oscillatory behavior of the pattern}. The last two panels $e)$ and $f)$ show the temporal evolution of the pattern formation emphasizing the amplitude chimera state. 

As a last comment, we want to point out that the formation of the chimera states is based on the assumption of the partial disorder in the eigenvector entries which for different structural reasons might not always be the case as can be noticed in the tenth eigenvector where a cluster of entries (yellow color) can be observed. Although such clusters might exist, the only neat cluster which is always present in the eigenvectors is that of the zero entries immediately related to the almost triangularity of the Laplacian matrix.

%%%%%%%%%%%%%%%%%%%%%%%%%%%%%%%%%%%%%%%%%%%%%%%%%%%%%%%%%%%%%%%%%%%%%%%%%%%%%%%%%%%%%%%%%%%%%%%%
%%%%%%%%%%%%%%%%%%%%%%%%%%%%%%%%%%%%%%%%%%%%%%%%%%%%%%%%%%%%%%%%%%%%%%%%%%%%%%%%%%%%%%%%%%%%%%%%

\section{Hybrid chimera patterns: the oscillon states}
\label{sec:III}
\noindent

Chimera states have been traditionally related to models of coupled dynamical systems where each node represents an oscillator characterized by an intrinsic phase variable to which, as in our case, might also be associated with the amplitude variable. Nevertheless, very recently an alternative mechanism was developed to show that chimera states and cluster synchronization can also emerge in system where the nodes, when uncoupled, are in a fixed point rather than limit cycle regime \cite{symm_break}. Following a symmetry-breaking mechanism as the one described in this paper, the authors of Ref. \cite{symm_break}, showed that due to an oscillatory instability, namely where the eigenvalues of the extended Jacobian have both real and imaginary parts, not only can global oscillations occur, but they can also self-organize in clusters of coherent or incoherent oscillatory patterns. Recalling that no oscillatory dynamics was imposed at the node level, the cause for the emergence of such behavior is found on the global coupling which is due either to the directedness of the network considered or a minimum of 3 observables (e.g., species) per node can generate the coherent-incoherent oscillations. In \cite{symm_break}, a key ingredient to have clusters of similar and different entries in the Laplacian eigenvectors was to consider modular networks which have the peculiarity of having eigenvectors with clusters of similar entries corresponding to the modules' nodes. As it can be intuited, such a role in non-normal networks is played by the zero entries of the eigenvectors, however, with a striking difference as we will discuss in the following.

%\onecolumngrid

\begin{figure*}%[h!]
	\centering
	\includegraphics[width = \linewidth]{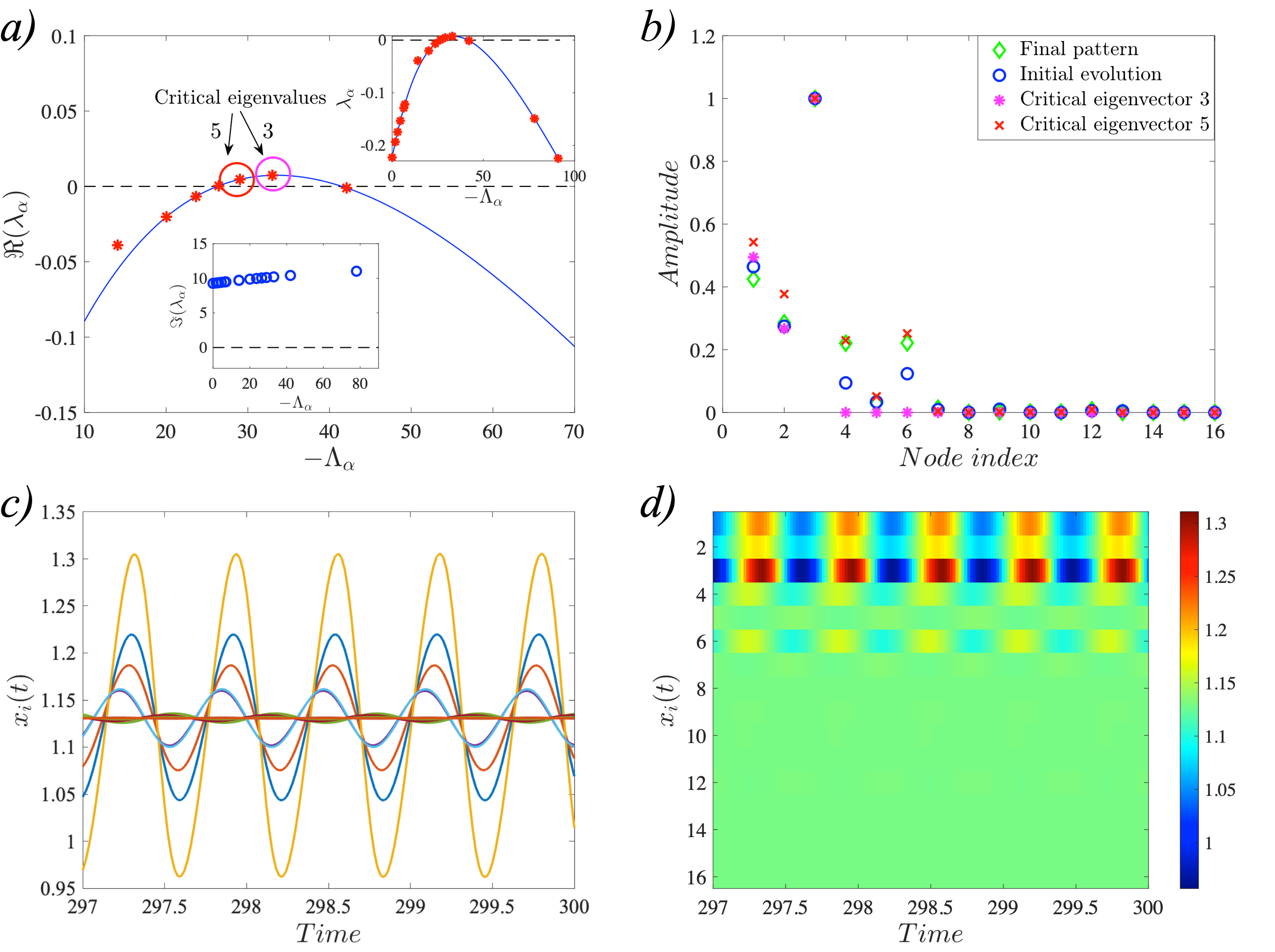}
	\caption{\textbf{Emergence of oscillon patterns.} $\textbf{a)}$ The Master Stability Function (dispersion relation) real (main, red stars) and imaginary (inset, blue circles) parts for the \textit{macaques competition} network \cite{ref41}. The critical eigenvalues correspond to the {third and the fifth igenvectors} shown in Fig. 1 $b)$. $\textbf{b)}$ Comparison between the critical {eigenvectors 3 (magenta stars) and 5 (red crosses), respectively,} and the normalized amplitudes of the initial evolution (blue circles) and the final pattern (green diamonds). $\textbf{c)}$ The time series for each nodes where it can be noticed that for the nodes from $7$ to $16$, (almost) no oscillations are present. $\textbf{d)}$ The colormap representation of the evolution of the oscillon patterns. The colorbar quantifies the oscillators amplitudes. The parameters for the Zhabotinsky model are $c_1=c_3=28.5$, $c_2=c_4=15.5$, $c_5=c_6=1$, $c_7=25.65$, $c_8=3.1$, $D_x=D_y=0$ and $D_z=0.1$.}
	\label{Fig2}
\end{figure*}

%\twocolumngrid

In this section, we present the emergence of oscillon patterns in a network support which consists of oscillations localized in a subset of nodes while the rest have a uniform stationary state \cite{Vanag_Epstein, oscillon_net}. The definition of such patterns, observed initially in granular media experiments, is extended to the network domain. These patterns consist of an oscillatory localized pattern surrounded by a stationary homogeneous neighborhood \cite{umbanhowar_localized_1996}. With this aim, we need to consider a 3 species model which in our case is described by the following set of equations:  
\begin{equation}
\begin{cases}
f(x_i,y_i,z_i)=-c_1x_iy_i^2  + c_3z_i^2 - c_7x_i/(q+x_i)\\
g(x_i,y_i,z_i)=c_2u_iy_i^2 - c_4y_i + c_8\\
h(x_i,y_i,z_i)=c_5x_i - c_6z_i
\end{cases}
\label{eq:Zhab}
\end{equation} 
introduced by Zhabotinsky \textit{et al.}, \cite{zhab, asllani_linear_2013}, where $q=10^{-4}$ and the rest of the constants $c_1-c_8$ represent parameters. Let us notice also that the reason why we are considering a multiple species model for the oscillatory instability rather than obtaining it solely from the directedness of the network \cite{asllani_theory_2014}, is due to empirical evidence. Real networks with strong non-normality have been observed to have a significantly small (or sometimes even absent) imaginary part in their spectrum compared to the real part \cite{asllani2018structure}. The latter has been shown analytically and numerically to contribute to a lesser extent to the emergence of global oscillations \cite{asllani_theory_2014, Asllani2018PRE, Muolo2021}. Starting from these premises in Fig. \ref{Fig2} panel $a)$ the MSF of the Zhabotinsky model in the \emph{macaques competition} network \cite{ref41} is presented, where in the inset the imaginary part of the extended Jacobian eigenvalues is displayed. We recall that in this case {the Jacobian matrix is time-independent} making the linearized dynamical system autonomous and the Master Stability Function reduces to the well-known dispersion relation \cite{Murray2008}. Consequently, the spectrum of the Jacobian can be obtained analytically through solving the eigenvalue problem \eqref{eq:master}. In panel $b)$ it is possible to appreciate the result of this approach where the critical eigenvectors corresponding to the unstable eigenvalues shown in the previous panel, are plotted versus the initial and final patterns. The temporal evolution of the oscillon pattern is presented in the consecutive panels $c)$ and $d)$. Following the amplification of the Laplacian eigenvector entries, several nodes will jump from their original position at the fixed point and start oscillating. {Notice that in this case the final pattern is jointly shaped by the third and fifth eigenvectors.} However, most of the nodes of the network in this case will remain in an (almost) stationary regime by keeping their original state at the fixed point. In summary, the network will exhibit an oscillon state, where localized nodes exhibit oscillatory behavior while the surrounding nodes remain stationary.

%%%%%%%%%%%%%%%%%%%%%%%%%%%%%%%%%%%%%%%%%%%%%%%%%%%%%%%%%%%%%%%%%%%%%%%%%%%%%%%%%%%%%%%%%%%%%%%%
%%%%%%%%%%%%%%%%%%%%%%%%%%%%%%%%%%%%%%%%%%%%%%%%%%%%%%%%%%%%%%%%%%%%%%%%%%%%%%%%%%%%%%%%%%%%%%%%

{\section{Are phase chimeras ubiquitous in the real world or is global synchronization elusive?}}
\label{sec:IV}

Another prominent example of coherent-incoherent patterns is that of phase chimera states which historically precedes the amplitude states discussed in the previous section \cite{chimera, Abrams_Strogatz}. {In the context of non-normal networks, the explanation for their emergence is} based not only on a different mechanism but also a local structural property (compared to the \emph{global} non-normality discussed earlier), that of the leader nodes which is ubiquitous for {empirical} networks. This time we will consider the \emph{ants dominance} network \cite{ref44}, which is made of a total of $16$ nodes (individuals) $9$ of which are source (leader) nodes indicated in red in Fig. \ref{Fig2} $a)$. For the Brusselator model set in the oscillator regime, we select parameters for which the system is strongly stable \footnote{Hereby, by ``strongly stable'' we mean that the stability indicators as the Jacobian eigenvalue with the largest real part, respectively, the Maximum Lyapunov Exponent have a considerably large magnitude apart from being negative.}, panel $b)$. Despite the algebraic degeneracy in the spectrum (the Laplacian has $9$ repeated zeros), the MSF formalism can still be applied due to a full basis of eigenvectors. However, if we perturb the network of coupled Brusselators with some moderate perturbations, we notice that the time series presented in panels $c)$ and $d)$ display a clear (phase) chimera behavior where although all nodes share the same exact amplitude, the oscillators corresponding to the source ones are out of sync compared to the rest of network (except for node $2$) which is very well synchronized. 
%\onecolumngrid

\begin{figure*}%[h!]
	\centering
	\includegraphics[width = \linewidth]{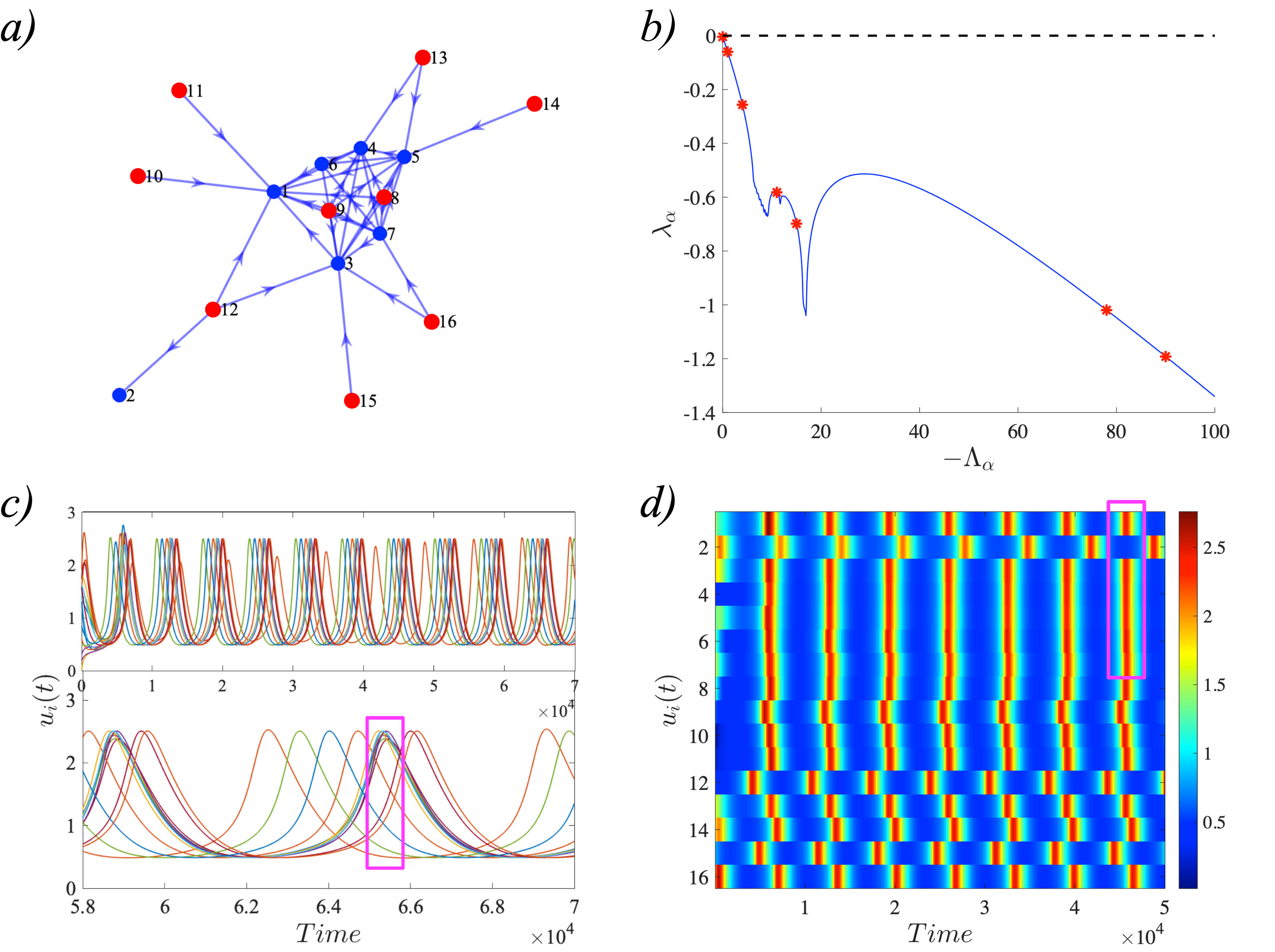}
	\caption{\textbf{Emergence of phase chimera states.} $\textbf{a)}$ The graphic representation of the \textit{ants dominance} network \cite{ref44} ordered hierarchically. $\textbf{b)}$ The Master Stability Function shows no (local) instability, however, there are $9$ zero overlaying eigenvalues in the MSF curve.  $\textbf{c)}$ The time series for each oscillator zoomed (lower part) and the complete evolution (upper part). $\textbf{d)}$ The colormap representation of the oscillators dynamics evolution with the synchronized clusters emphasized with the magenta rectangle. The colorbar quantifies the oscillators amplitudes. {Notice that due to the random initial perturbation of the system, some of the sources belong to the synchronized cluster just by chance.} The parameters for the Brusselator model are $b=2.5$, $c=1$, $D_u=0.0175$, and $D_v=0.075$.}
	\label{Fig2}
\end{figure*}

%\twocolumngrid
The observed behavior can be attributed to the presence of a source node that is disconnected from the rest of the network in terms of incoming links. When the system is perturbed and the source node is initially triggered, it eventually settles back to its original limit cycle. The settling phase might differ from the initial phase which was the same for all the network. On the other hand, such set of nodes will act as forcing terms in our dynamical system and consequently are expected to influence the rest of the group. This is for instance the case for node $2$ which is also a leader node, but a sink rather than a source one. It is immediately ``forced'' by the source node $12$ which in combination with the intrinsic phase of node $2$ yields as a result a different final phase. However, this is not always the case, since the set of nodes forced by the source leaders might form a densely connected cluster with each other resulting in disturbances coming from the source nodes being inhibited or averaged across the cluster. This is the case for the nodes 1 and 3-7, which form a robust synchronized cluster. Consequently the entire network splits in two groups of coherent and incoherent oscillators resulting in a classic phase chimera state. As a final note, it is worth noting that although the system under consideration is not asymptotically stable, it still exhibits Lyapunov stability as far as the orbits of the perturbed system will stay forever near the initial equilibrium point \cite{strogatz_book}. 

The scenario of phase chimeras illustrated in this section might not always be the case. In fact, the different phase impulses arriving from the source nodes can keep the rest of the network away from a synchronized state. The role of source nodes, similarly to the mechanism described above, has also been recently studied in Ref. \cite{wright_central_2019} in term of synchronization robustness, although no mention was made related to a chimera-like phenomenon. Based on the bowtie architecture of many complex networks, the authors raise the question that synchronization might be difficult to achieve in practice. However, the question we address in this paper is not about the mechanism itself, which is intuitively straightforward. Instead, our focus is on the prevalence of leader source nodes in real networks and the implications this has for the emergence of coherent-incoherent patterns. {Also an important difference with \cite{wright_central_2019} is that our results exclude the presence of any synchronized source cluster on the network since source SCC do not exist when leaders are present \footnote{{In principle, a SCC connected with the rest of the network with only outgoing links, can in principle synchronize as shown in \cite{symm_break} and consequently decrease the disordered oscillations of the leaders.}}.} This empirical evidence, together with the fact that the regime of parameters for which this behavior occurs is considerably large, i.e., the only requirement is that the system should be stable, unavoidably poses the question if chimera states are indeed much more present in real-world systems than initially anticipated. The recent trend of experimental studies focused on chimera states further strengthen this assertion \cite{chimera_brain, chimera_firefly}. {Another notable conclusion that arises from this analysis is the significant challenge associated with achieving global synchronization in real networks. Instead, the attainable level of synchronization is often limited to local subnetworks, manifesting as a chimera state. However,} it is also worthmentioning that such conclusions are obtained from the current state of art models of synchronization dynamics and that there is need for more sophisticated ones to better understand such phenomenon.

%%%%%%%%%%%%%%%%%%%%%%%%%%%%%%%%%%%%%%%%%%%%%%%%%%%%%%%%%%%%%%%%%%%%%%%%%%%%%%%%%%%%%%%%%%%%%%%%
%%%%%%%%%%%%%%%%%%%%%%%%%%%%%%%%%%%%%%%%%%%%%%%%%%%%%%%%%%%%%%%%%%%%%%%%%%%%%%%%%%%%%%%%%%%%%%%%

\section{Discussion and conclusions}
\label{sec:V}
\noindent 

Most natural systems are characterized by complex interacting structures which collectively play a crucial role in their system dynamics. Aiming to understand common structural features of the interaction networks, it was recently discovered that a vast majority of them manifest a strong non-normality \cite{asllani2018structure}. From the structural point of view, this implies that the majority of real networks exhibit characteristics akin to directed acyclic graphs (DAGs), thus possessing a strong hierarchical topology \cite{asllani2018structure, OBrien_NN}. As a consequence, from a mathematical perspective, adjacency matrices and related operators representing such networks, are close to being triangular where perfectly triangular corresponds to a DAG network. In other words, the eigenvectors of {their matrix operators} have patterns where many entries are zero, and the non-zero entries are generally scattered randomly without a specific structure. Based on this remarkable characteristic of real world networks, in this paper we have presented a systematic mechanism for the generation of chimera patterns. We prove that the spontaneous emergence of amplitude chimera states can be explained by a symmetry-breaking mechanism according to which the final patterns resemble the structure of the eigenvectors of the unstable mode(s) selected from the synchronized manifold. 

{While synchronization generally requires individual (coupled) dynamical systems to exhibit an oscillatory behavior, recent studies have shown that it is still possible to obtain synchronized-desynchronized collective oscillations by destabilizing a homogeneous fixed point \cite{symm_break}. In this novel scenario, a three-variable model is a prerequisite for generating an oscillatory instability responsible for the emergence of global oscillations. In the case of non-normal networks,} a hybrid chimera pattern arises, where unsynchronized oscillations coexist with stationary nodes. We have named such patterns oscillons due to being reminiscent to the localized and isolated oscillations that emerge in granular media \cite{Vanag_Epstein}.

Non-normal networks observed in nature, have shown to possess another distinct feature, the presence of nodes that are sources (or sinks), named as leader nodes \cite{OBrien_NN}. The peculiarity is their omnipresence in non-normal networks and the fact that no other strongly connected components coexist in the same network. {In the context of synchronization, source nodes serve as forcing terms that have the potential to disrupt synchronization at any level within a subnetwork. Consequently, leader (source) nodes pose a constant threat to the dynamics of synchronization. Building upon this empirical observation, we demonstrate that regardless of the chosen parameters, a stable system of coupled oscillators can still exhibit phase chimera states. This occurs because source nodes, being disconnected from the influence of the rest of the network, retain the phase shift resulting from the initial perturbation. Meanwhile, the remaining nodes may potentially return to a synchronized regime, effectively absorbing perturbations originating from the source nodes.} 

In conclusion, in this work our contribution is multifold. Firstly, we present a unique approach to explain the emergence of several patterns such as amplitude and phase chimeras also as oscillon states all sharing the common feature of coexistence of coherent and incoherent states. This mechanism is grounded on the spectral properties characterizing real-world networks. Secondly, we put to the fore the problem that the presence of non-normality in empirical networks poses a persistent challenge to understanding their synchronization behavior, which indicates the need for more advanced models and approaches to comprehensively grasp and explain these phenomena.

\section*{Acknowledgements}
The work of R.M. is supported by a FRIA-FNRS PhD fellowship, grant FC 33443, funded by the Walloon region. R.M. and J.D.O'B. acknowledge funding from the Bridge Grant of the yrCSS. This work was partly supported by Science Foundation Ireland under Grant number 16/IA/4470.

%\appendix

\section*{Methods}
\label{sec:AppA}
\noindent
\begin{theorem}
The right eigenvectors of a triangular matrix $\textbf{A}_{n\times n}$ form a triangular matrix $\textbf{P}_{n\times n}$ when considered as columns of $\textbf{P}$ and ordered according to the eigenvalues of $\textbf{A}$. 
\end{theorem}

\underline{\textit{Proof}:} We will prove the results by considering that the vector with the first entry non-zero is an eigenvector of the first eigenvalue, the vectors with the first two entries non-zero is an eigenvector of the second eigenvalue and so on. So if we consider the matrix $\textbf{A}$ with entries $a_{i,j}$ with $i\leq j$ then the vector $\mathbf{v}_1= \left[1,0,\dots,0 \right]^T$ is the eigenvector corresponding to the eigenvalue $a_{1,1}$ (recall that the diagonal entries are the eigenvalues of $\textbf{A}$). In fact,  
\begin{equation*}
\begin{cases}
\cancel{\left(a_{1,1}-a_{1,1}\right)}\times 1+a_{1,2}\times 0+\dots+a_{1,n}\times 0=0\\
\left(a_{i,i}-a_{1,1}\right)\times 0+a_{i,i+1}\times 0+\dots+ a_{i,n}\times 0=0, \;\;\; i>1\,.
\end{cases}
\end{equation*}
For a general eigenvector with the first $m\leq n$ entries non-zero $\mathbf{v}_m= \left[x_1,\dots, x_m, 0,\dots,0 \right]^T$ we have
\begin{equation*}
\left(\mathbf{A}_{m\times m}-a_{m,m}\mathbf{I}_{m\times m}\right)\mathbf{v}_m=\mathbf{0}
\end{equation*}
where $\mathbf{A}_{m\times m}$ is the triangular block matrix made from the first $m$ rows and the first $m$ columns of $\textbf{A}$. Clearly since at least one eigenvalue, i.e., diagonal entry of the matrix $\mathbf{A}_{m\times m}-a_{m,m}\mathbf{I}_{m\times m}$ is zero, the determinant is also zero and thus the system always has a non trivial solution $\mathbf{v}_m$. If instead of the eigenvalue $a_{m,m}$ we would have chosen, $a_{i,i}, i\neq m$ then we would have had $\left(a_{m,m}-a_{i,i}\right)x_m=0$ and thus $x_m=0$ if $a_{m,m}\neq a_{i,i}$. This would then correspond to consider the previous eigenvector $\mathbf{v}_{m-1}$. %Notice also that in case we have repeated eigenvalues

%%%%%%%%%% Insert bibliography here %%%%%%%%%%%%%%

\bibliography{NN_bib}

\begin{thebibliography}{115}%
\makeatletter
\providecommand \@ifxundefined [1]{%
 \@ifx{#1\undefined}
}%
\providecommand \@ifnum [1]{%
 \ifnum #1\expandafter \@firstoftwo
 \else \expandafter \@secondoftwo
 \fi
}%
\providecommand \@ifx [1]{%
 \ifx #1\expandafter \@firstoftwo
 \else \expandafter \@secondoftwo
 \fi
}%
\providecommand \natexlab [1]{#1}%
\providecommand \enquote  [1]{``#1''}%
\providecommand \bibnamefont  [1]{#1}%
\providecommand \bibfnamefont [1]{#1}%
\providecommand \citenamefont [1]{#1}%
\providecommand \href@noop [0]{\@secondoftwo}%
\providecommand \href [0]{\begingroup \@sanitize@url \@href}%
\providecommand \@href[1]{\@@startlink{#1}\@@href}%
\providecommand \@@href[1]{\endgroup#1\@@endlink}%
\providecommand \@sanitize@url [0]{\catcode `\\12\catcode `\$12\catcode
  `\&12\catcode `\#12\catcode `\^12\catcode `\_12\catcode `\%12\relax}%
\providecommand \@@startlink[1]{}%
\providecommand \@@endlink[0]{}%
\providecommand \url  [0]{\begingroup\@sanitize@url \@url }%
\providecommand \@url [1]{\endgroup\@href {#1}{\urlprefix }}%
\providecommand \urlprefix  [0]{URL }%
\providecommand \Eprint [0]{\href }%
\providecommand \doibase [0]{https://doi.org/}%
\providecommand \selectlanguage [0]{\@gobble}%
\providecommand \bibinfo  [0]{\@secondoftwo}%
\providecommand \bibfield  [0]{\@secondoftwo}%
\providecommand \translation [1]{[#1]}%
\providecommand \BibitemOpen [0]{}%
\providecommand \bibitemStop [0]{}%
\providecommand \bibitemNoStop [0]{.\EOS\space}%
\providecommand \EOS [0]{\spacefactor3000\relax}%
\providecommand \BibitemShut  [1]{\csname bibitem#1\endcsname}%
\let\auto@bib@innerbib\@empty
%</preamble>
\bibitem [{\citenamefont {Simon}(1991)}]{Simon1991}%
  \BibitemOpen
  \bibfield  {author} {\bibinfo {author} {\bibfnamefont {H.~A.}\ \bibnamefont
  {Simon}},\ }\bibinfo {title} {The architecture of complexity},\ in\ \href
  {https://doi.org/10.1007/978-1-4899-0718-9_31} {\emph {\bibinfo {booktitle}
  {Facets of Systems Science}}}\ (\bibinfo  {publisher} {Springer US},\
  \bibinfo {address} {Boston, MA},\ \bibinfo {year} {1991})\ pp.\ \bibinfo
  {pages} {457--476}\BibitemShut {NoStop}%
\bibitem [{\citenamefont {Newman}(2010)}]{newman2010networks}%
  \BibitemOpen
  \bibfield  {author} {\bibinfo {author} {\bibfnamefont {M.}~\bibnamefont
  {Newman}},\ }\href
  {https://doi.org/10.1093/acprof:oso/9780199206650.001.0001} {\emph {\bibinfo
  {title} {Networks: An Introduction}}}\ (\bibinfo  {publisher} {Oxford
  university press},\ \bibinfo {year} {2010})\BibitemShut {NoStop}%
\bibitem [{\citenamefont {Pikovskij}\ \emph {et~al.}(2007)\citenamefont
  {Pikovskij}, \citenamefont {Rosenblum},\ and\ \citenamefont
  {Kurths}}]{pikovskij_synchronization:_2007}%
  \BibitemOpen
  \bibfield  {author} {\bibinfo {author} {\bibfnamefont {A.}~\bibnamefont
  {Pikovskij}}, \bibinfo {author} {\bibfnamefont {M.}~\bibnamefont
  {Rosenblum}},\ and\ \bibinfo {author} {\bibfnamefont {J.}~\bibnamefont
  {Kurths}},\ }\href@noop {} {\emph {\bibinfo {title} {Synchronization: a
  universal concept in nonlinear sciences}}}\ (\bibinfo  {publisher} {Cambridge
  Univ. Press},\ \bibinfo {address} {Cambridge},\ \bibinfo {year}
  {2007})\BibitemShut {NoStop}%
\bibitem [{\citenamefont {Arenas}\ \emph {et~al.}(2008)\citenamefont {Arenas},
  \citenamefont {Díaz-Guilera}, \citenamefont {Kurths}, \citenamefont
  {Moreno},\ and\ \citenamefont {Zhou}}]{arenas_synchronization_2008}%
  \BibitemOpen
  \bibfield  {author} {\bibinfo {author} {\bibfnamefont {A.}~\bibnamefont
  {Arenas}}, \bibinfo {author} {\bibfnamefont {A.}~\bibnamefont
  {Díaz-Guilera}}, \bibinfo {author} {\bibfnamefont {J.}~\bibnamefont
  {Kurths}}, \bibinfo {author} {\bibfnamefont {Y.}~\bibnamefont {Moreno}},\
  and\ \bibinfo {author} {\bibfnamefont {C.}~\bibnamefont {Zhou}},\ }\href
  {https://doi.org/10.1016/j.physrep.2008.09.002} {\bibfield  {journal}
  {\bibinfo  {journal} {Physics Reports}\ }\textbf {\bibinfo {volume} {469}},\
  \bibinfo {pages} {93} (\bibinfo {year} {2008})}\BibitemShut {NoStop}%
\bibitem [{\citenamefont {Chouzouris}\ \emph {et~al.}(2018)\citenamefont
  {Chouzouris}, \citenamefont {Omelchenko}, \citenamefont {Zakharova},
  \citenamefont {Hlinka}, \citenamefont {Jiruska},\ and\ \citenamefont
  {Schöll}}]{Chouzouris}%
  \BibitemOpen
  \bibfield  {author} {\bibinfo {author} {\bibfnamefont {T.}~\bibnamefont
  {Chouzouris}}, \bibinfo {author} {\bibfnamefont {I.}~\bibnamefont
  {Omelchenko}}, \bibinfo {author} {\bibfnamefont {A.}~\bibnamefont
  {Zakharova}}, \bibinfo {author} {\bibfnamefont {J.}~\bibnamefont {Hlinka}},
  \bibinfo {author} {\bibfnamefont {P.}~\bibnamefont {Jiruska}},\ and\ \bibinfo
  {author} {\bibfnamefont {E.}~\bibnamefont {Schöll}},\ }\href
  {https://doi.org/10.1063/1.5009812} {\bibfield  {journal} {\bibinfo
  {journal} {Chaos: An Interdisciplinary Journal of Nonlinear Science}\
  }\textbf {\bibinfo {volume} {28}},\ \bibinfo {pages} {045112} (\bibinfo
  {year} {2018})},\ \Eprint
  {https://arxiv.org/abs/https://doi.org/10.1063/1.5009812}
  {https://doi.org/10.1063/1.5009812} \BibitemShut {NoStop}%
\bibitem [{\citenamefont {Buck}\ and\ \citenamefont
  {Buck}(1976)}]{john_synchronous_1976}%
  \BibitemOpen
  \bibfield  {author} {\bibinfo {author} {\bibfnamefont {J.}~\bibnamefont
  {Buck}}\ and\ \bibinfo {author} {\bibfnamefont {E.}~\bibnamefont {Buck}},\
  }\href {https://doi.org/10.1038/scientificamerican0576-74} {\bibfield
  {journal} {\bibinfo  {journal} {Scientific American}\ }\textbf {\bibinfo
  {volume} {234}},\ \bibinfo {pages} {74} (\bibinfo {year} {1976})}\BibitemShut
  {NoStop}%
\bibitem [{\citenamefont {Motter}\ \emph {et~al.}(2013)\citenamefont {Motter},
  \citenamefont {Myers}, \citenamefont {Anghel},\ and\ \citenamefont
  {Nishikawa}}]{motter_spontaneous_2013}%
  \BibitemOpen
  \bibfield  {author} {\bibinfo {author} {\bibfnamefont {A.~E.}\ \bibnamefont
  {Motter}}, \bibinfo {author} {\bibfnamefont {S.~A.}\ \bibnamefont {Myers}},
  \bibinfo {author} {\bibfnamefont {M.}~\bibnamefont {Anghel}},\ and\ \bibinfo
  {author} {\bibfnamefont {T.}~\bibnamefont {Nishikawa}},\ }\href
  {https://doi.org/10.1038/nphys2535} {\bibfield  {journal} {\bibinfo
  {journal} {Nature Physics}\ }\textbf {\bibinfo {volume} {9}},\ \bibinfo
  {pages} {191} (\bibinfo {year} {2013})}\BibitemShut {NoStop}%
\bibitem [{\citenamefont {Sivrikaya}\ and\ \citenamefont
  {Yener}(2004)}]{Sivrikaya_Yener}%
  \BibitemOpen
  \bibfield  {author} {\bibinfo {author} {\bibfnamefont {F.}~\bibnamefont
  {Sivrikaya}}\ and\ \bibinfo {author} {\bibfnamefont {B.}~\bibnamefont
  {Yener}},\ }\href {https://doi.org/10.1109/MNET.2004.1316761} {\bibfield
  {journal} {\bibinfo  {journal} {IEEE Network}\ }\textbf {\bibinfo {volume}
  {18}},\ \bibinfo {pages} {45} (\bibinfo {year} {2004})}\BibitemShut {NoStop}%
\bibitem [{\citenamefont {Kuramoto}(1975)}]{kuramoto}%
  \BibitemOpen
  \bibfield  {author} {\bibinfo {author} {\bibfnamefont {Y.}~\bibnamefont
  {Kuramoto}},\ }in\ \href@noop {} {\emph {\bibinfo {booktitle} {International
  Symposium on Mathematical Problems in Theoretical Physics}}},\ \bibinfo
  {editor} {edited by\ \bibinfo {editor} {\bibfnamefont {H.}~\bibnamefont
  {Araki}}}\ (\bibinfo  {publisher} {Springer Berlin Heidelberg},\ \bibinfo
  {address} {Berlin, Heidelberg},\ \bibinfo {year} {1975})\ pp.\ \bibinfo
  {pages} {420--422}\BibitemShut {NoStop}%
\bibitem [{\citenamefont {Kuramoto}(1984)}]{kuramoto_book}%
  \BibitemOpen
  \bibfield  {author} {\bibinfo {author} {\bibfnamefont {Y.}~\bibnamefont
  {Kuramoto}},\ }\href {https://doi.org/10.1007/978-3-642-69689-3} {\emph
  {\bibinfo {title} {Chemical {Oscillations}, {Waves}, and {Turbulence}}}},\
  edited by\ \bibinfo {editor} {\bibfnamefont {H.}~\bibnamefont {Haken}},\
  \bibinfo {series} {Springer {Series} in {Synergetics}}, Vol.~\bibinfo
  {volume} {19}\ (\bibinfo  {publisher} {Springer Berlin Heidelberg},\ \bibinfo
  {address} {Berlin, Heidelberg},\ \bibinfo {year} {1984})\BibitemShut
  {NoStop}%
\bibitem [{\citenamefont {Kuramoto}\ and\ \citenamefont
  {Battogtokh}(2002)}]{chimera}%
  \BibitemOpen
  \bibfield  {author} {\bibinfo {author} {\bibfnamefont {Y.}~\bibnamefont
  {Kuramoto}}\ and\ \bibinfo {author} {\bibfnamefont {D.}~\bibnamefont
  {Battogtokh}},\ }\href@noop {} {\bibfield  {journal} {\bibinfo  {journal}
  {Nonlinear Phenomena in Complex Systems}\ }\textbf {\bibinfo {volume} {5}},\
  \bibinfo {pages} {380 } (\bibinfo {year} {2002})}\BibitemShut {NoStop}%
\bibitem [{\citenamefont {Abrams}\ and\ \citenamefont
  {Strogatz}(2004)}]{Abrams_Strogatz}%
  \BibitemOpen
  \bibfield  {author} {\bibinfo {author} {\bibfnamefont {D.~M.}\ \bibnamefont
  {Abrams}}\ and\ \bibinfo {author} {\bibfnamefont {S.~H.}\ \bibnamefont
  {Strogatz}},\ }\href {https://doi.org/10.1103/PhysRevLett.93.174102}
  {\bibfield  {journal} {\bibinfo  {journal} {Phys. Rev. Lett.}\ }\textbf
  {\bibinfo {volume} {93}},\ \bibinfo {pages} {174102} (\bibinfo {year}
  {2004})}\BibitemShut {NoStop}%
\bibitem [{\citenamefont {Panaggio}\ and\ \citenamefont
  {Abrams}(2015)}]{Panaggio_2015}%
  \BibitemOpen
  \bibfield  {author} {\bibinfo {author} {\bibfnamefont {M.~J.}\ \bibnamefont
  {Panaggio}}\ and\ \bibinfo {author} {\bibfnamefont {D.~M.}\ \bibnamefont
  {Abrams}},\ }\href {https://doi.org/10.1088/0951-7715/28/3/R67} {\bibfield
  {journal} {\bibinfo  {journal} {Nonlinearity}\ }\textbf {\bibinfo {volume}
  {28}},\ \bibinfo {pages} {R67} (\bibinfo {year} {2015})}\BibitemShut
  {NoStop}%
\bibitem [{\citenamefont {Zakharova}(2020)}]{zakharova_chimera_2020}%
  \BibitemOpen
  \bibfield  {author} {\bibinfo {author} {\bibfnamefont {A.}~\bibnamefont
  {Zakharova}},\ }\href@noop {} {\emph {\bibinfo {title} {Chimera patterns in
  networks: interplay between dynamics, structure, noise, and delay}}},\
  Understanding complex systems\ (\bibinfo  {publisher} {Springer},\ \bibinfo
  {address} {Cham},\ \bibinfo {year} {2020})\BibitemShut {NoStop}%
\bibitem [{\citenamefont {Bansal}\ \emph {et~al.}(2019)\citenamefont {Bansal},
  \citenamefont {Garcia}, \citenamefont {Tompson}, \citenamefont {Verstynen},
  \citenamefont {Vettel},\ and\ \citenamefont {Muldoon}}]{chimera_brain}%
  \BibitemOpen
  \bibfield  {author} {\bibinfo {author} {\bibfnamefont {K.}~\bibnamefont
  {Bansal}}, \bibinfo {author} {\bibfnamefont {J.~O.}\ \bibnamefont {Garcia}},
  \bibinfo {author} {\bibfnamefont {S.~H.}\ \bibnamefont {Tompson}}, \bibinfo
  {author} {\bibfnamefont {T.}~\bibnamefont {Verstynen}}, \bibinfo {author}
  {\bibfnamefont {J.~M.}\ \bibnamefont {Vettel}},\ and\ \bibinfo {author}
  {\bibfnamefont {S.~F.}\ \bibnamefont {Muldoon}},\ }\href
  {https://doi.org/10.1126/sciadv.aau8535} {\bibfield  {journal} {\bibinfo
  {journal} {Science Advances}\ }\textbf {\bibinfo {volume} {5}},\ \bibinfo
  {pages} {eaau8535} (\bibinfo {year} {2019})},\ \Eprint
  {https://arxiv.org/abs/https://www.science.org/doi/pdf/10.1126/sciadv.aau8535}
  {https://www.science.org/doi/pdf/10.1126/sciadv.aau8535} \BibitemShut
  {NoStop}%
\bibitem [{\citenamefont {Sarfati}\ and\ \citenamefont
  {Peleg}(2022)}]{chimera_firefly}%
  \BibitemOpen
  \bibfield  {author} {\bibinfo {author} {\bibfnamefont {R.}~\bibnamefont
  {Sarfati}}\ and\ \bibinfo {author} {\bibfnamefont {O.}~\bibnamefont
  {Peleg}},\ }\href {https://doi.org/10.1126/sciadv.add6690} {\bibfield
  {journal} {\bibinfo  {journal} {Science Advances}\ }\textbf {\bibinfo
  {volume} {8}},\ \bibinfo {pages} {eadd6690} (\bibinfo {year} {2022})},\
  \Eprint
  {https://arxiv.org/abs/https://www.science.org/doi/pdf/10.1126/sciadv.add6690}
  {https://www.science.org/doi/pdf/10.1126/sciadv.add6690} \BibitemShut
  {NoStop}%
\bibitem [{\citenamefont {Trefethen}\ and\ \citenamefont
  {Embree}(2005)}]{Trefethen2005}%
  \BibitemOpen
  \bibfield  {author} {\bibinfo {author} {\bibfnamefont {L.~N.}\ \bibnamefont
  {Trefethen}}\ and\ \bibinfo {author} {\bibfnamefont {M.}~\bibnamefont
  {Embree}},\ }\href {https://doi.org/10.2307/j.ctvzxx9kj} {\emph {\bibinfo
  {title} {Spectra and Pseudospectra}}}\ (\bibinfo  {publisher} {Princeton
  University Press},\ \bibinfo {year} {2005})\BibitemShut {NoStop}%
\bibitem [{\citenamefont {Siebert}\ \emph {et~al.}(2020)\citenamefont
  {Siebert}, \citenamefont {Hall}, \citenamefont {Gleeson},\ and\ \citenamefont
  {Asllani}}]{Siebert}%
  \BibitemOpen
  \bibfield  {author} {\bibinfo {author} {\bibfnamefont {B.~A.}\ \bibnamefont
  {Siebert}}, \bibinfo {author} {\bibfnamefont {C.~L.}\ \bibnamefont {Hall}},
  \bibinfo {author} {\bibfnamefont {J.~P.}\ \bibnamefont {Gleeson}},\ and\
  \bibinfo {author} {\bibfnamefont {M.}~\bibnamefont {Asllani}},\ }\href
  {https://doi.org/10.1103/PhysRevE.102.052306} {\bibfield  {journal} {\bibinfo
   {journal} {Phys. Rev. E}\ }\textbf {\bibinfo {volume} {102}},\ \bibinfo
  {pages} {052306} (\bibinfo {year} {2020})}\BibitemShut {NoStop}%
\bibitem [{\citenamefont {Asllani}\ \emph {et~al.}(2022)\citenamefont
  {Asllani}, \citenamefont {Siebert}, \citenamefont {Arenas},\ and\
  \citenamefont {Gleeson}}]{symm_break}%
  \BibitemOpen
  \bibfield  {author} {\bibinfo {author} {\bibfnamefont {M.}~\bibnamefont
  {Asllani}}, \bibinfo {author} {\bibfnamefont {B.~A.}\ \bibnamefont
  {Siebert}}, \bibinfo {author} {\bibfnamefont {A.}~\bibnamefont {Arenas}},\
  and\ \bibinfo {author} {\bibfnamefont {J.~P.}\ \bibnamefont {Gleeson}},\
  }\href {https://doi.org/10.1063/5.0060466} {\bibfield  {journal} {\bibinfo
  {journal} {Chaos: An Interdisciplinary Journal of Nonlinear Science}\
  }\textbf {\bibinfo {volume} {32}},\ \bibinfo {pages} {013107} (\bibinfo
  {year} {2022})},\ \Eprint
  {https://arxiv.org/abs/https://doi.org/10.1063/5.0060466}
  {https://doi.org/10.1063/5.0060466} \BibitemShut {NoStop}%
\bibitem [{\citenamefont {Asllani}\ \emph {et~al.}(2018)\citenamefont
  {Asllani}, \citenamefont {Lambiotte},\ and\ \citenamefont
  {Carletti}}]{asllani2018structure}%
  \BibitemOpen
  \bibfield  {author} {\bibinfo {author} {\bibfnamefont {M.}~\bibnamefont
  {Asllani}}, \bibinfo {author} {\bibfnamefont {R.}~\bibnamefont {Lambiotte}},\
  and\ \bibinfo {author} {\bibfnamefont {T.}~\bibnamefont {Carletti}},\ }\href
  {https://doi.org/10.1126/sciadv.aau9403} {\bibfield  {journal} {\bibinfo
  {journal} {Sci. Adv.}\ }\textbf {\bibinfo {volume} {4}},\ \bibinfo {pages}
  {eaau9403} (\bibinfo {year} {2018})}\BibitemShut {NoStop}%
\bibitem [{\citenamefont {O'Brien}\ \emph {et~al.}(2021)\citenamefont
  {O'Brien}, \citenamefont {Oliveira}, \citenamefont {Gleeson},\ and\
  \citenamefont {Asllani}}]{OBrien_NN}%
  \BibitemOpen
  \bibfield  {author} {\bibinfo {author} {\bibfnamefont {J.~D.}\ \bibnamefont
  {O'Brien}}, \bibinfo {author} {\bibfnamefont {K.~A.}\ \bibnamefont
  {Oliveira}}, \bibinfo {author} {\bibfnamefont {J.~P.}\ \bibnamefont
  {Gleeson}},\ and\ \bibinfo {author} {\bibfnamefont {M.}~\bibnamefont
  {Asllani}},\ }\href {https://doi.org/10.1103/PhysRevResearch.3.023117}
  {\bibfield  {journal} {\bibinfo  {journal} {Phys. Rev. Res.}\ }\textbf
  {\bibinfo {volume} {3}},\ \bibinfo {pages} {023117} (\bibinfo {year}
  {2021})}\BibitemShut {NoStop}%
\bibitem [{\citenamefont {Cowell}\ \emph {et~al.}(2007)\citenamefont {Cowell},
  \citenamefont {Blake},\ and\ \citenamefont {Russell}}]{ref68}%
  \BibitemOpen
  \bibfield  {author} {\bibinfo {author} {\bibfnamefont {R.~M.}\ \bibnamefont
  {Cowell}}, \bibinfo {author} {\bibfnamefont {K.~R.}\ \bibnamefont {Blake}},\
  and\ \bibinfo {author} {\bibfnamefont {J.~W.}\ \bibnamefont {Russell}},\
  }\href@noop {} {\bibfield  {journal} {\bibinfo  {journal} {Journal of
  Comparative Neurology}\ }\textbf {\bibinfo {volume} {502}},\ \bibinfo {pages}
  {1} (\bibinfo {year} {2007})}\BibitemShut {NoStop}%
\bibitem [{\citenamefont {Harriger}\ \emph {et~al.}(2012)\citenamefont
  {Harriger}, \citenamefont {Van Den~Heuvel},\ and\ \citenamefont
  {Sporns}}]{ref69}%
  \BibitemOpen
  \bibfield  {author} {\bibinfo {author} {\bibfnamefont {L.}~\bibnamefont
  {Harriger}}, \bibinfo {author} {\bibfnamefont {M.~P.}\ \bibnamefont {Van
  Den~Heuvel}},\ and\ \bibinfo {author} {\bibfnamefont {O.}~\bibnamefont
  {Sporns}},\ }\href@noop {} {\bibfield  {journal} {\bibinfo  {journal} {PloS
  one}\ }\textbf {\bibinfo {volume} {7}},\ \bibinfo {pages} {e46497} (\bibinfo
  {year} {2012})}\BibitemShut {NoStop}%
\bibitem [{\citenamefont {Markov}\ \emph {et~al.}(2013)\citenamefont {Markov},
  \citenamefont {Ercsey-Ravasz}, \citenamefont {Lamy}, \citenamefont {Gomes},
  \citenamefont {Magrou}, \citenamefont {Misery}, \citenamefont {Giroud},
  \citenamefont {Barone}, \citenamefont {Dehay}, \citenamefont {Toroczkai}
  \emph {et~al.}}]{ref70}%
  \BibitemOpen
  \bibfield  {author} {\bibinfo {author} {\bibfnamefont {N.~T.}\ \bibnamefont
  {Markov}}, \bibinfo {author} {\bibfnamefont {M.}~\bibnamefont
  {Ercsey-Ravasz}}, \bibinfo {author} {\bibfnamefont {C.}~\bibnamefont {Lamy}},
  \bibinfo {author} {\bibfnamefont {A.~R.~R.}\ \bibnamefont {Gomes}}, \bibinfo
  {author} {\bibfnamefont {L.}~\bibnamefont {Magrou}}, \bibinfo {author}
  {\bibfnamefont {P.}~\bibnamefont {Misery}}, \bibinfo {author} {\bibfnamefont
  {P.}~\bibnamefont {Giroud}}, \bibinfo {author} {\bibfnamefont
  {P.}~\bibnamefont {Barone}}, \bibinfo {author} {\bibfnamefont
  {C.}~\bibnamefont {Dehay}}, \bibinfo {author} {\bibfnamefont
  {Z.}~\bibnamefont {Toroczkai}}, \emph {et~al.},\ }\href
  {https://doi.org/10.1073/pnas.1218972110} {\bibfield  {journal} {\bibinfo
  {journal} {Proc. Natl. Acad. Sci. USA}\ }\textbf {\bibinfo {volume} {110}},\
  \bibinfo {pages} {5187} (\bibinfo {year} {2013})}\BibitemShut {NoStop}%
\bibitem [{\citenamefont {Kaiser}\ and\ \citenamefont
  {Hilgetag}(2006)}]{Kaiser2006}%
  \BibitemOpen
  \bibfield  {author} {\bibinfo {author} {\bibfnamefont {M.}~\bibnamefont
  {Kaiser}}\ and\ \bibinfo {author} {\bibfnamefont {C.~C.}\ \bibnamefont
  {Hilgetag}},\ }\bibfield  {journal} {\bibinfo  {journal} {PLoS Computational
  Biology}\ }\href {https://doi.org/10.1371/journal.pcbi.0020095}
  {10.1371/journal.pcbi.0020095} (\bibinfo {year} {2006})\BibitemShut {NoStop}%
\bibitem [{\citenamefont {Watts}\ and\ \citenamefont {Strogatz}(1998)}]{ref54}%
  \BibitemOpen
  \bibfield  {author} {\bibinfo {author} {\bibfnamefont {D.~J.}\ \bibnamefont
  {Watts}}\ and\ \bibinfo {author} {\bibfnamefont {S.~H.}\ \bibnamefont
  {Strogatz}},\ }\href {https://doi.org/10.1038/30918} {\bibfield  {journal}
  {\bibinfo  {journal} {Nature}\ }\textbf {\bibinfo {volume} {393}},\ \bibinfo
  {pages} {440} (\bibinfo {year} {1998})}\BibitemShut {NoStop}%
\bibitem [{\citenamefont {Bota}\ and\ \citenamefont {Swanson}(2007)}]{ref66}%
  \BibitemOpen
  \bibfield  {author} {\bibinfo {author} {\bibfnamefont {M.}~\bibnamefont
  {Bota}}\ and\ \bibinfo {author} {\bibfnamefont {L.~W.}\ \bibnamefont
  {Swanson}},\ }\href@noop {} {\bibfield  {journal} {\bibinfo  {journal}
  {Journal of Comparative Neurology}\ }\textbf {\bibinfo {volume} {500}},\
  \bibinfo {pages} {807} (\bibinfo {year} {2007})}\BibitemShut {NoStop}%
\bibitem [{\citenamefont {Carere}\ \emph {et~al.}(2007)\citenamefont {Carere},
  \citenamefont {Ball},\ and\ \citenamefont {Balthazart}}]{ref67}%
  \BibitemOpen
  \bibfield  {author} {\bibinfo {author} {\bibfnamefont {C.}~\bibnamefont
  {Carere}}, \bibinfo {author} {\bibfnamefont {G.~F.}\ \bibnamefont {Ball}},\
  and\ \bibinfo {author} {\bibfnamefont {J.}~\bibnamefont {Balthazart}},\
  }\href@noop {} {\bibfield  {journal} {\bibinfo  {journal} {Journal of
  Comparative Neurology}\ }\textbf {\bibinfo {volume} {500}},\ \bibinfo {pages}
  {894} (\bibinfo {year} {2007})}\BibitemShut {NoStop}%
\bibitem [{\citenamefont {Johnson}()}]{ref28}%
  \BibitemOpen
  \bibfield  {author} {\bibinfo {author} {\bibfnamefont {S.}~\bibnamefont
  {Johnson}},\ }\href@noop {} {\bibinfo {title} {Network data repository from
  various sources.}},\ \bibinfo {howpublished}
  {\url{https://www.samuel-johnson.org/data}}\BibitemShut {NoStop}%
\bibitem [{\citenamefont {de~Reus}\ and\ \citenamefont {van~den
  Heuvel}(2013)}]{ref43}%
  \BibitemOpen
  \bibfield  {author} {\bibinfo {author} {\bibfnamefont {M.~A.}\ \bibnamefont
  {de~Reus}}\ and\ \bibinfo {author} {\bibfnamefont {M.~P.}\ \bibnamefont
  {van~den Heuvel}},\ }\href@noop {} {\bibfield  {journal} {\bibinfo  {journal}
  {Journal of Neuroscience}\ }\textbf {\bibinfo {volume} {33}},\ \bibinfo
  {pages} {12929} (\bibinfo {year} {2013})}\BibitemShut {NoStop}%
\bibitem [{\citenamefont {Thieffry}\ \emph {et~al.}(1998)\citenamefont
  {Thieffry}, \citenamefont {Huerta}, \citenamefont {P{\'e}rez-Rueda},\ and\
  \citenamefont {Collado-Vides}}]{ref55}%
  \BibitemOpen
  \bibfield  {author} {\bibinfo {author} {\bibfnamefont {D.}~\bibnamefont
  {Thieffry}}, \bibinfo {author} {\bibfnamefont {A.~M.}\ \bibnamefont
  {Huerta}}, \bibinfo {author} {\bibfnamefont {E.}~\bibnamefont
  {P{\'e}rez-Rueda}},\ and\ \bibinfo {author} {\bibfnamefont {J.}~\bibnamefont
  {Collado-Vides}},\ }\href
  {https://doi.org/10.1002/(SICI)1521-1878(199805)20:5<433::AID-BIES10>3.0.CO;2-2}
  {\bibfield  {journal} {\bibinfo  {journal} {Bioessays}\ }\textbf {\bibinfo
  {volume} {20}},\ \bibinfo {pages} {433} (\bibinfo {year} {1998})}\BibitemShut
  {NoStop}%
\bibitem [{\citenamefont {Gerstein}\ \emph {et~al.}(2012)\citenamefont
  {Gerstein}, \citenamefont {Kundaje}, \citenamefont {Hariharan}, \citenamefont
  {Landt}, \citenamefont {Yan}, \citenamefont {Cheng}, \citenamefont {Mu},
  \citenamefont {Khurana}, \citenamefont {Rozowsky}, \citenamefont {Alexander},
  \citenamefont {Min}, \citenamefont {Alves}, \citenamefont {Abyzov},
  \citenamefont {Addleman}, \citenamefont {Bhardwaj}, \citenamefont {Boyle},
  \citenamefont {Cayting}, \citenamefont {Charos}, \citenamefont {Chen},
  \citenamefont {Cheng}, \citenamefont {Clarke}, \citenamefont {Eastman},
  \citenamefont {Euskirchen}, \citenamefont {Frietze}, \citenamefont {Fu},
  \citenamefont {Gertz}, \citenamefont {Grubert}, \citenamefont {Harmanci},
  \citenamefont {Jain}, \citenamefont {Kasowski}, \citenamefont {Lacroute},
  \citenamefont {Leng}, \citenamefont {Lian}, \citenamefont {Monahan},
  \citenamefont {Oĝgeen}, \citenamefont {Ouyang}, \citenamefont {Partridge},
  \citenamefont {Patacsil}, \citenamefont {Pauli}, \citenamefont {Raha},
  \citenamefont {Ramirez}, \citenamefont {Reddy}, \citenamefont {Reed},
  \citenamefont {Shi}, \citenamefont {Slifer}, \citenamefont {Wang},
  \citenamefont {Wu}, \citenamefont {Yang}, \citenamefont {Yip}, \citenamefont
  {Zilberman-Schapira}, \citenamefont {Batzoglou}, \citenamefont {Sidow},
  \citenamefont {Farnham}, \citenamefont {Myers}, \citenamefont {Weissman},\
  and\ \citenamefont {Snyder}}]{Gerstein2012}%
  \BibitemOpen
  \bibfield  {author} {\bibinfo {author} {\bibfnamefont {M.~B.}\ \bibnamefont
  {Gerstein}}, \bibinfo {author} {\bibfnamefont {A.}~\bibnamefont {Kundaje}},
  \bibinfo {author} {\bibfnamefont {M.}~\bibnamefont {Hariharan}}, \bibinfo
  {author} {\bibfnamefont {S.~G.}\ \bibnamefont {Landt}}, \bibinfo {author}
  {\bibfnamefont {K.~K.}\ \bibnamefont {Yan}}, \bibinfo {author} {\bibfnamefont
  {C.}~\bibnamefont {Cheng}}, \bibinfo {author} {\bibfnamefont {X.~J.}\
  \bibnamefont {Mu}}, \bibinfo {author} {\bibfnamefont {E.}~\bibnamefont
  {Khurana}}, \bibinfo {author} {\bibfnamefont {J.}~\bibnamefont {Rozowsky}},
  \bibinfo {author} {\bibfnamefont {R.}~\bibnamefont {Alexander}}, \bibinfo
  {author} {\bibfnamefont {R.}~\bibnamefont {Min}}, \bibinfo {author}
  {\bibfnamefont {P.}~\bibnamefont {Alves}}, \bibinfo {author} {\bibfnamefont
  {A.}~\bibnamefont {Abyzov}}, \bibinfo {author} {\bibfnamefont
  {N.}~\bibnamefont {Addleman}}, \bibinfo {author} {\bibfnamefont
  {N.}~\bibnamefont {Bhardwaj}}, \bibinfo {author} {\bibfnamefont {A.~P.}\
  \bibnamefont {Boyle}}, \bibinfo {author} {\bibfnamefont {P.}~\bibnamefont
  {Cayting}}, \bibinfo {author} {\bibfnamefont {A.}~\bibnamefont {Charos}},
  \bibinfo {author} {\bibfnamefont {D.~Z.}\ \bibnamefont {Chen}}, \bibinfo
  {author} {\bibfnamefont {Y.}~\bibnamefont {Cheng}}, \bibinfo {author}
  {\bibfnamefont {D.}~\bibnamefont {Clarke}}, \bibinfo {author} {\bibfnamefont
  {C.}~\bibnamefont {Eastman}}, \bibinfo {author} {\bibfnamefont
  {G.}~\bibnamefont {Euskirchen}}, \bibinfo {author} {\bibfnamefont
  {S.}~\bibnamefont {Frietze}}, \bibinfo {author} {\bibfnamefont
  {Y.}~\bibnamefont {Fu}}, \bibinfo {author} {\bibfnamefont {J.}~\bibnamefont
  {Gertz}}, \bibinfo {author} {\bibfnamefont {F.}~\bibnamefont {Grubert}},
  \bibinfo {author} {\bibfnamefont {A.}~\bibnamefont {Harmanci}}, \bibinfo
  {author} {\bibfnamefont {P.}~\bibnamefont {Jain}}, \bibinfo {author}
  {\bibfnamefont {M.}~\bibnamefont {Kasowski}}, \bibinfo {author}
  {\bibfnamefont {P.}~\bibnamefont {Lacroute}}, \bibinfo {author}
  {\bibfnamefont {J.}~\bibnamefont {Leng}}, \bibinfo {author} {\bibfnamefont
  {J.}~\bibnamefont {Lian}}, \bibinfo {author} {\bibfnamefont {H.}~\bibnamefont
  {Monahan}}, \bibinfo {author} {\bibfnamefont {H.}~\bibnamefont {Oĝgeen}},
  \bibinfo {author} {\bibfnamefont {Z.}~\bibnamefont {Ouyang}}, \bibinfo
  {author} {\bibfnamefont {E.~C.}\ \bibnamefont {Partridge}}, \bibinfo {author}
  {\bibfnamefont {D.}~\bibnamefont {Patacsil}}, \bibinfo {author}
  {\bibfnamefont {F.}~\bibnamefont {Pauli}}, \bibinfo {author} {\bibfnamefont
  {D.}~\bibnamefont {Raha}}, \bibinfo {author} {\bibfnamefont {L.}~\bibnamefont
  {Ramirez}}, \bibinfo {author} {\bibfnamefont {T.~E.}\ \bibnamefont {Reddy}},
  \bibinfo {author} {\bibfnamefont {B.}~\bibnamefont {Reed}}, \bibinfo {author}
  {\bibfnamefont {M.}~\bibnamefont {Shi}}, \bibinfo {author} {\bibfnamefont
  {T.}~\bibnamefont {Slifer}}, \bibinfo {author} {\bibfnamefont
  {J.}~\bibnamefont {Wang}}, \bibinfo {author} {\bibfnamefont {L.}~\bibnamefont
  {Wu}}, \bibinfo {author} {\bibfnamefont {X.}~\bibnamefont {Yang}}, \bibinfo
  {author} {\bibfnamefont {K.~Y.}\ \bibnamefont {Yip}}, \bibinfo {author}
  {\bibfnamefont {G.}~\bibnamefont {Zilberman-Schapira}}, \bibinfo {author}
  {\bibfnamefont {S.}~\bibnamefont {Batzoglou}}, \bibinfo {author}
  {\bibfnamefont {A.}~\bibnamefont {Sidow}}, \bibinfo {author} {\bibfnamefont
  {P.~J.}\ \bibnamefont {Farnham}}, \bibinfo {author} {\bibfnamefont {R.~M.}\
  \bibnamefont {Myers}}, \bibinfo {author} {\bibfnamefont {S.~M.}\ \bibnamefont
  {Weissman}},\ and\ \bibinfo {author} {\bibfnamefont {M.}~\bibnamefont
  {Snyder}},\ }\bibfield  {journal} {\bibinfo  {journal} {Nature}\ }\href
  {https://doi.org/10.1038/nature11245} {10.1038/nature11245} (\bibinfo {year}
  {2012})\BibitemShut {NoStop}%
\bibitem [{\citenamefont {Harbison}\ \emph {et~al.}(2004)\citenamefont
  {Harbison}, \citenamefont {Gordon}, \citenamefont {Lee}, \citenamefont
  {Rinaldi}, \citenamefont {Macisaac}, \citenamefont {Danford}, \citenamefont
  {Hannett}, \citenamefont {Tagne}, \citenamefont {Reynolds}, \citenamefont
  {Yoo}, \citenamefont {Jennings}, \citenamefont {Zeitlinger}, \citenamefont
  {Pokholok}, \citenamefont {Kellis}, \citenamefont {Rolfe}, \citenamefont
  {Takusagawa}, \citenamefont {Lander}, \citenamefont {Gifford}, \citenamefont
  {Fraenkel},\ and\ \citenamefont {Young}}]{ref59}%
  \BibitemOpen
  \bibfield  {author} {\bibinfo {author} {\bibfnamefont {C.~T.}\ \bibnamefont
  {Harbison}}, \bibinfo {author} {\bibfnamefont {D.~B.}\ \bibnamefont
  {Gordon}}, \bibinfo {author} {\bibfnamefont {T.~I.}\ \bibnamefont {Lee}},
  \bibinfo {author} {\bibfnamefont {N.~K.}\ \bibnamefont {Rinaldi}}, \bibinfo
  {author} {\bibfnamefont {K.~D.}\ \bibnamefont {Macisaac}}, \bibinfo {author}
  {\bibfnamefont {T.~W.}\ \bibnamefont {Danford}}, \bibinfo {author}
  {\bibfnamefont {N.~M.}\ \bibnamefont {Hannett}}, \bibinfo {author}
  {\bibfnamefont {J.-B.}\ \bibnamefont {Tagne}}, \bibinfo {author}
  {\bibfnamefont {D.~B.}\ \bibnamefont {Reynolds}}, \bibinfo {author}
  {\bibfnamefont {J.}~\bibnamefont {Yoo}}, \bibinfo {author} {\bibfnamefont
  {E.~G.}\ \bibnamefont {Jennings}}, \bibinfo {author} {\bibfnamefont
  {J.}~\bibnamefont {Zeitlinger}}, \bibinfo {author} {\bibfnamefont {D.~K.}\
  \bibnamefont {Pokholok}}, \bibinfo {author} {\bibfnamefont {M.}~\bibnamefont
  {Kellis}}, \bibinfo {author} {\bibfnamefont {P.~A.}\ \bibnamefont {Rolfe}},
  \bibinfo {author} {\bibfnamefont {K.~T.}\ \bibnamefont {Takusagawa}},
  \bibinfo {author} {\bibfnamefont {E.~S.}\ \bibnamefont {Lander}}, \bibinfo
  {author} {\bibfnamefont {D.~K.}\ \bibnamefont {Gifford}}, \bibinfo {author}
  {\bibfnamefont {E.}~\bibnamefont {Fraenkel}},\ and\ \bibinfo {author}
  {\bibfnamefont {R.~A.}\ \bibnamefont {Young}},\ }\href
  {https://doi.org/10.1038/nature02800} {\bibfield  {journal} {\bibinfo
  {journal} {Nature}\ }\textbf {\bibinfo {volume} {431}},\ \bibinfo {pages}
  {99} (\bibinfo {year} {2004})}\BibitemShut {NoStop}%
\bibitem [{\citenamefont {Costanzo}\ \emph {et~al.}(2001)\citenamefont
  {Costanzo}, \citenamefont {Crawford}, \citenamefont {Hirschman},
  \citenamefont {Kranz}, \citenamefont {Olsen}, \citenamefont {Robertson},
  \citenamefont {Skrzypek}, \citenamefont {Braun}, \citenamefont {Hopkins},
  \citenamefont {Kondu} \emph {et~al.}}]{ref62}%
  \BibitemOpen
  \bibfield  {author} {\bibinfo {author} {\bibfnamefont {M.~C.}\ \bibnamefont
  {Costanzo}}, \bibinfo {author} {\bibfnamefont {M.~E.}\ \bibnamefont
  {Crawford}}, \bibinfo {author} {\bibfnamefont {J.~E.}\ \bibnamefont
  {Hirschman}}, \bibinfo {author} {\bibfnamefont {J.~E.}\ \bibnamefont
  {Kranz}}, \bibinfo {author} {\bibfnamefont {P.}~\bibnamefont {Olsen}},
  \bibinfo {author} {\bibfnamefont {L.~S.}\ \bibnamefont {Robertson}}, \bibinfo
  {author} {\bibfnamefont {M.~S.}\ \bibnamefont {Skrzypek}}, \bibinfo {author}
  {\bibfnamefont {B.~R.}\ \bibnamefont {Braun}}, \bibinfo {author}
  {\bibfnamefont {K.~L.}\ \bibnamefont {Hopkins}}, \bibinfo {author}
  {\bibfnamefont {P.}~\bibnamefont {Kondu}}, \emph {et~al.},\ }\href
  {https://doi.org/10.1093/nar/29.1.75} {\bibfield  {journal} {\bibinfo
  {journal} {Nucleic Acids Research}\ }\textbf {\bibinfo {volume} {29}},\
  \bibinfo {pages} {75} (\bibinfo {year} {2001})}\BibitemShut {NoStop}%
\bibitem [{\citenamefont {Sanz}\ \emph {et~al.}(2011)\citenamefont {Sanz},
  \citenamefont {Navarro}, \citenamefont {Arbu{\'{e}}s}, \citenamefont
  {Mart{\'{i}}n}, \citenamefont {Mariju{\'{a}}n},\ and\ \citenamefont
  {Moreno}}]{Sanz2011}%
  \BibitemOpen
  \bibfield  {author} {\bibinfo {author} {\bibfnamefont {J.}~\bibnamefont
  {Sanz}}, \bibinfo {author} {\bibfnamefont {J.}~\bibnamefont {Navarro}},
  \bibinfo {author} {\bibfnamefont {A.}~\bibnamefont {Arbu{\'{e}}s}}, \bibinfo
  {author} {\bibfnamefont {C.}~\bibnamefont {Mart{\'{i}}n}}, \bibinfo {author}
  {\bibfnamefont {P.~C.}\ \bibnamefont {Mariju{\'{a}}n}},\ and\ \bibinfo
  {author} {\bibfnamefont {Y.}~\bibnamefont {Moreno}},\ }\bibfield  {journal}
  {\bibinfo  {journal} {PLoS ONE}\ }\href
  {https://doi.org/10.1371/journal.pone.0022178} {10.1371/journal.pone.0022178}
  (\bibinfo {year} {2011})\BibitemShut {NoStop}%
\bibitem [{\citenamefont {Jeong}\ \emph {et~al.}(2000)\citenamefont {Jeong},
  \citenamefont {Tombor}, \citenamefont {Albert}, \citenamefont {Oltvai},\ and\
  \citenamefont {Barab{\'a}si}}]{ref52}%
  \BibitemOpen
  \bibfield  {author} {\bibinfo {author} {\bibfnamefont {H.}~\bibnamefont
  {Jeong}}, \bibinfo {author} {\bibfnamefont {B.}~\bibnamefont {Tombor}},
  \bibinfo {author} {\bibfnamefont {R.}~\bibnamefont {Albert}}, \bibinfo
  {author} {\bibfnamefont {Z.~N.}\ \bibnamefont {Oltvai}},\ and\ \bibinfo
  {author} {\bibfnamefont {A.-L.}\ \bibnamefont {Barab{\'a}si}},\ }\href
  {https://doi.org/10.1038/35036627} {\bibfield  {journal} {\bibinfo  {journal}
  {Nature}\ }\textbf {\bibinfo {volume} {407}},\ \bibinfo {pages} {651}
  (\bibinfo {year} {2000})}\BibitemShut {NoStop}%
\bibitem [{\citenamefont {Ewing}\ \emph {et~al.}(2007)\citenamefont {Ewing},
  \citenamefont {Chu}, \citenamefont {Elisma}, \citenamefont {Li},
  \citenamefont {Taylor}, \citenamefont {Climie}, \citenamefont
  {McBroom-Cerajewski}, \citenamefont {Robinson}, \citenamefont {O'Connor},
  \citenamefont {Li}, \citenamefont {Taylor}, \citenamefont {Dharsee},
  \citenamefont {Ho}, \citenamefont {Heilbut}, \citenamefont {Moore},
  \citenamefont {Zhang}, \citenamefont {Ornatsky}, \citenamefont {Bukhman},
  \citenamefont {Ethier}, \citenamefont {Sheng}, \citenamefont {Vasilescu},
  \citenamefont {Abu-Farha}, \citenamefont {Lambert}, \citenamefont {Duewel},
  \citenamefont {Stewart}, \citenamefont {Kuehl}, \citenamefont {Hogue},
  \citenamefont {Colwill}, \citenamefont {Gladwish}, \citenamefont {Muskat},
  \citenamefont {Kinach}, \citenamefont {Adams}, \citenamefont {Moran},
  \citenamefont {Morin}, \citenamefont {Topaloglou},\ and\ \citenamefont
  {Figeys}}]{Ewing2007}%
  \BibitemOpen
  \bibfield  {author} {\bibinfo {author} {\bibfnamefont {R.~M.}\ \bibnamefont
  {Ewing}}, \bibinfo {author} {\bibfnamefont {P.}~\bibnamefont {Chu}}, \bibinfo
  {author} {\bibfnamefont {F.}~\bibnamefont {Elisma}}, \bibinfo {author}
  {\bibfnamefont {H.}~\bibnamefont {Li}}, \bibinfo {author} {\bibfnamefont
  {P.}~\bibnamefont {Taylor}}, \bibinfo {author} {\bibfnamefont
  {S.}~\bibnamefont {Climie}}, \bibinfo {author} {\bibfnamefont
  {L.}~\bibnamefont {McBroom-Cerajewski}}, \bibinfo {author} {\bibfnamefont
  {M.~D.}\ \bibnamefont {Robinson}}, \bibinfo {author} {\bibfnamefont
  {L.}~\bibnamefont {O'Connor}}, \bibinfo {author} {\bibfnamefont
  {M.}~\bibnamefont {Li}}, \bibinfo {author} {\bibfnamefont {R.}~\bibnamefont
  {Taylor}}, \bibinfo {author} {\bibfnamefont {M.}~\bibnamefont {Dharsee}},
  \bibinfo {author} {\bibfnamefont {Y.}~\bibnamefont {Ho}}, \bibinfo {author}
  {\bibfnamefont {A.}~\bibnamefont {Heilbut}}, \bibinfo {author} {\bibfnamefont
  {L.}~\bibnamefont {Moore}}, \bibinfo {author} {\bibfnamefont
  {S.}~\bibnamefont {Zhang}}, \bibinfo {author} {\bibfnamefont
  {O.}~\bibnamefont {Ornatsky}}, \bibinfo {author} {\bibfnamefont {Y.~V.}\
  \bibnamefont {Bukhman}}, \bibinfo {author} {\bibfnamefont {M.}~\bibnamefont
  {Ethier}}, \bibinfo {author} {\bibfnamefont {Y.}~\bibnamefont {Sheng}},
  \bibinfo {author} {\bibfnamefont {J.}~\bibnamefont {Vasilescu}}, \bibinfo
  {author} {\bibfnamefont {M.}~\bibnamefont {Abu-Farha}}, \bibinfo {author}
  {\bibfnamefont {J.~P.}\ \bibnamefont {Lambert}}, \bibinfo {author}
  {\bibfnamefont {H.~S.}\ \bibnamefont {Duewel}}, \bibinfo {author}
  {\bibfnamefont {I.~I.}\ \bibnamefont {Stewart}}, \bibinfo {author}
  {\bibfnamefont {B.}~\bibnamefont {Kuehl}}, \bibinfo {author} {\bibfnamefont
  {K.}~\bibnamefont {Hogue}}, \bibinfo {author} {\bibfnamefont
  {K.}~\bibnamefont {Colwill}}, \bibinfo {author} {\bibfnamefont
  {K.}~\bibnamefont {Gladwish}}, \bibinfo {author} {\bibfnamefont
  {B.}~\bibnamefont {Muskat}}, \bibinfo {author} {\bibfnamefont
  {R.}~\bibnamefont {Kinach}}, \bibinfo {author} {\bibfnamefont {S.~L.}\
  \bibnamefont {Adams}}, \bibinfo {author} {\bibfnamefont {M.~F.}\ \bibnamefont
  {Moran}}, \bibinfo {author} {\bibfnamefont {G.~B.}\ \bibnamefont {Morin}},
  \bibinfo {author} {\bibfnamefont {T.}~\bibnamefont {Topaloglou}},\ and\
  \bibinfo {author} {\bibfnamefont {D.}~\bibnamefont {Figeys}},\ }\bibfield
  {journal} {\bibinfo  {journal} {Molecular Systems Biology}\ }\href
  {https://doi.org/10.1038/msb4100134} {10.1038/msb4100134} (\bibinfo {year}
  {2007})\BibitemShut {NoStop}%
\bibitem [{\citenamefont {De~Nooy}(1999)}]{ref35}%
  \BibitemOpen
  \bibfield  {author} {\bibinfo {author} {\bibfnamefont {W.}~\bibnamefont
  {De~Nooy}},\ }\href@noop {} {\bibfield  {journal} {\bibinfo  {journal}
  {Poetics}\ }\textbf {\bibinfo {volume} {26}},\ \bibinfo {pages} {385}
  (\bibinfo {year} {1999})}\BibitemShut {NoStop}%
\bibitem [{\citenamefont {da~Cunha}\ \emph {et~al.}(2020)\citenamefont
  {da~Cunha}, \citenamefont {MacCarron}, \citenamefont {Passold}, \citenamefont
  {dos Santos}, \citenamefont {Oliveira},\ and\ \citenamefont
  {Gleeson}}]{ref27}%
  \BibitemOpen
  \bibfield  {author} {\bibinfo {author} {\bibfnamefont {B.~R.}\ \bibnamefont
  {da~Cunha}}, \bibinfo {author} {\bibfnamefont {P.}~\bibnamefont {MacCarron}},
  \bibinfo {author} {\bibfnamefont {J.~F.}\ \bibnamefont {Passold}}, \bibinfo
  {author} {\bibfnamefont {L.~W.}\ \bibnamefont {dos Santos}}, \bibinfo
  {author} {\bibfnamefont {K.~A.}\ \bibnamefont {Oliveira}},\ and\ \bibinfo
  {author} {\bibfnamefont {J.~P.}\ \bibnamefont {Gleeson}},\ }\href
  {https://doi.org/10.1038/s41598-019-56704-4} {\bibfield  {journal} {\bibinfo
  {journal} {Scientific Reports}\ }\textbf {\bibinfo {volume} {10}},\ \bibinfo
  {pages} {73} (\bibinfo {year} {2020})}\BibitemShut {NoStop}%
\bibitem [{\citenamefont {Linvill}\ and\ \citenamefont {Warren}(2020)}]{ref72}%
  \BibitemOpen
  \bibfield  {author} {\bibinfo {author} {\bibfnamefont {D.~L.}\ \bibnamefont
  {Linvill}}\ and\ \bibinfo {author} {\bibfnamefont {P.~L.}\ \bibnamefont
  {Warren}},\ }\href {https://doi.org/10.1080/10584609.2020.1718257} {\bibfield
   {journal} {\bibinfo  {journal} {Political Communication}\ }\textbf {\bibinfo
  {volume} {37}},\ \bibinfo {pages} {447} (\bibinfo {year} {2020})},\ \Eprint
  {https://arxiv.org/abs/https://doi.org/10.1080/10584609.2020.1718257}
  {https://doi.org/10.1080/10584609.2020.1718257} \BibitemShut {NoStop}%
\bibitem [{\citenamefont {Leskovec}\ \emph {et~al.}(2010)\citenamefont
  {Leskovec}, \citenamefont {Huttenlocher},\ and\ \citenamefont
  {Kleinberg}}]{ref18}%
  \BibitemOpen
  \bibfield  {author} {\bibinfo {author} {\bibfnamefont {J.}~\bibnamefont
  {Leskovec}}, \bibinfo {author} {\bibfnamefont {D.}~\bibnamefont
  {Huttenlocher}},\ and\ \bibinfo {author} {\bibfnamefont {J.}~\bibnamefont
  {Kleinberg}},\ }in\ \href@noop {} {\emph {\bibinfo {booktitle} {Proceedings
  of the SIGCHI conference on human factors in computing systems}}}\ (\bibinfo
  {year} {2010})\ pp.\ \bibinfo {pages} {1361--1370}\BibitemShut {NoStop}%
\bibitem [{\citenamefont {Adamic}\ and\ \citenamefont {Glance}(2005)}]{ref1}%
  \BibitemOpen
  \bibfield  {author} {\bibinfo {author} {\bibfnamefont {L.~A.}\ \bibnamefont
  {Adamic}}\ and\ \bibinfo {author} {\bibfnamefont {N.}~\bibnamefont
  {Glance}},\ }in\ \href@noop {} {\emph {\bibinfo {booktitle} {Proceedings of
  the 3rd international workshop on Link discovery}}}\ (\bibinfo {year}
  {2005})\ pp.\ \bibinfo {pages} {36--43}\BibitemShut {NoStop}%
\bibitem [{\citenamefont {Kaggle}()}]{ref25}%
  \BibitemOpen
  \bibfield  {author} {\bibinfo {author} {\bibnamefont {Kaggle}},\ }\href@noop
  {} {\bibinfo {title} {Chess ratings - elo versus the rest of the world
  (2010).}},\ \bibinfo {howpublished}
  {\url{https://www.kaggle.com/c/chess/data}}\BibitemShut {NoStop}%
\bibitem [{\citenamefont {Coleman}\ \emph {et~al.}(1957)\citenamefont
  {Coleman}, \citenamefont {Katz},\ and\ \citenamefont {Menzel}}]{ref8}%
  \BibitemOpen
  \bibfield  {author} {\bibinfo {author} {\bibfnamefont {J.}~\bibnamefont
  {Coleman}}, \bibinfo {author} {\bibfnamefont {E.}~\bibnamefont {Katz}},\ and\
  \bibinfo {author} {\bibfnamefont {H.}~\bibnamefont {Menzel}},\ }\href@noop {}
  {\bibfield  {journal} {\bibinfo  {journal} {Sociometry}\ }\textbf {\bibinfo
  {volume} {20}},\ \bibinfo {pages} {253} (\bibinfo {year} {1957})}\BibitemShut
  {NoStop}%
\bibitem [{\citenamefont {{J. Kunegis}}()}]{konect}%
  \BibitemOpen
  \bibfield  {author} {\bibinfo {author} {\bibnamefont {{J. Kunegis}}},\
  }\href@noop {} {\bibinfo {title} {{The KONECT project}}},\ \bibinfo
  {howpublished} {\url{http://konect.cc/}}\BibitemShut {NoStop}%
\bibitem [{\citenamefont {Opsahl}\ and\ \citenamefont
  {Panzarasa}(2009)}]{Opsahl2009}%
  \BibitemOpen
  \bibfield  {author} {\bibinfo {author} {\bibfnamefont {T.}~\bibnamefont
  {Opsahl}}\ and\ \bibinfo {author} {\bibfnamefont {P.}~\bibnamefont
  {Panzarasa}},\ }\bibfield  {journal} {\bibinfo  {journal} {Social Networks}\
  }\href {https://doi.org/10.1016/j.socnet.2009.02.002}
  {10.1016/j.socnet.2009.02.002} (\bibinfo {year} {2009})\BibitemShut {NoStop}%
\bibitem [{\citenamefont {Thompson}\ and\ \citenamefont
  {Townsend}(2003)}]{Thompson2003}%
  \BibitemOpen
  \bibfield  {author} {\bibinfo {author} {\bibfnamefont {R.~M.}\ \bibnamefont
  {Thompson}}\ and\ \bibinfo {author} {\bibfnamefont {C.~R.}\ \bibnamefont
  {Townsend}},\ }\bibfield  {journal} {\bibinfo  {journal} {Ecology}\ }\href
  {https://doi.org/10.1890/0012-9658(2003)084[0145:IOSFWO]2.0.CO;2}
  {10.1890/0012-9658(2003)084[0145:IOSFWO]2.0.CO;2} (\bibinfo {year}
  {2003})\BibitemShut {NoStop}%
\bibitem [{\citenamefont {Thompson}\ and\ \citenamefont
  {Mcintosh}(1998)}]{ref5}%
  \BibitemOpen
  \bibfield  {author} {\bibinfo {author} {\bibfnamefont {R.~M.}\ \bibnamefont
  {Thompson}}\ and\ \bibinfo {author} {\bibfnamefont {A.~R.}\ \bibnamefont
  {Mcintosh}},\ }\href@noop {} {\bibfield  {journal} {\bibinfo  {journal}
  {Ecology Letters}\ }\textbf {\bibinfo {volume} {1}},\ \bibinfo {pages} {200}
  (\bibinfo {year} {1998})}\BibitemShut {NoStop}%
\bibitem [{\citenamefont {Klaise}\ and\ \citenamefont {Johnson}(2017)}]{ref9}%
  \BibitemOpen
  \bibfield  {author} {\bibinfo {author} {\bibfnamefont {J.}~\bibnamefont
  {Klaise}}\ and\ \bibinfo {author} {\bibfnamefont {S.}~\bibnamefont
  {Johnson}},\ }\href@noop {} {\bibfield  {journal} {\bibinfo  {journal}
  {Scientific reports}\ }\textbf {\bibinfo {volume} {7}},\ \bibinfo {pages} {1}
  (\bibinfo {year} {2017})}\BibitemShut {NoStop}%
\bibitem [{\citenamefont {Dunne}\ \emph {et~al.}(2013)\citenamefont {Dunne},
  \citenamefont {Lafferty}, \citenamefont {Dobson}, \citenamefont {Hechinger},
  \citenamefont {Kuris}, \citenamefont {Martinez}, \citenamefont {McLaughlin},
  \citenamefont {Mouritsen}, \citenamefont {Poulin}, \citenamefont {Reise}
  \emph {et~al.}}]{ref2}%
  \BibitemOpen
  \bibfield  {author} {\bibinfo {author} {\bibfnamefont {J.~A.}\ \bibnamefont
  {Dunne}}, \bibinfo {author} {\bibfnamefont {K.~D.}\ \bibnamefont {Lafferty}},
  \bibinfo {author} {\bibfnamefont {A.~P.}\ \bibnamefont {Dobson}}, \bibinfo
  {author} {\bibfnamefont {R.~F.}\ \bibnamefont {Hechinger}}, \bibinfo {author}
  {\bibfnamefont {A.~M.}\ \bibnamefont {Kuris}}, \bibinfo {author}
  {\bibfnamefont {N.~D.}\ \bibnamefont {Martinez}}, \bibinfo {author}
  {\bibfnamefont {J.~P.}\ \bibnamefont {McLaughlin}}, \bibinfo {author}
  {\bibfnamefont {K.~N.}\ \bibnamefont {Mouritsen}}, \bibinfo {author}
  {\bibfnamefont {R.}~\bibnamefont {Poulin}}, \bibinfo {author} {\bibfnamefont
  {K.}~\bibnamefont {Reise}}, \emph {et~al.},\ }\href@noop {} {\bibfield
  {journal} {\bibinfo  {journal} {PLoS Biol}\ }\textbf {\bibinfo {volume}
  {11}},\ \bibinfo {pages} {e1001579} (\bibinfo {year} {2013})}\BibitemShut
  {NoStop}%
\bibitem [{\citenamefont {Thompson}\ and\ \citenamefont
  {Townsend}(2005)}]{ref13}%
  \BibitemOpen
  \bibfield  {author} {\bibinfo {author} {\bibfnamefont {R.~M.}\ \bibnamefont
  {Thompson}}\ and\ \bibinfo {author} {\bibfnamefont {C.}~\bibnamefont
  {Townsend}},\ }\href@noop {} {\bibfield  {journal} {\bibinfo  {journal}
  {Oikos}\ }\textbf {\bibinfo {volume} {108}},\ \bibinfo {pages} {137}
  (\bibinfo {year} {2005})}\BibitemShut {NoStop}%
\bibitem [{\citenamefont {Memmott}\ \emph {et~al.}(2000)\citenamefont
  {Memmott}, \citenamefont {Martinez},\ and\ \citenamefont {Cohen}}]{ref22}%
  \BibitemOpen
  \bibfield  {author} {\bibinfo {author} {\bibfnamefont {J.}~\bibnamefont
  {Memmott}}, \bibinfo {author} {\bibfnamefont {N.~D.}\ \bibnamefont
  {Martinez}},\ and\ \bibinfo {author} {\bibfnamefont {J.}~\bibnamefont
  {Cohen}},\ }\href@noop {} {\bibfield  {journal} {\bibinfo  {journal} {Journal
  of Animal Ecology}\ }\textbf {\bibinfo {volume} {69}},\ \bibinfo {pages} {1}
  (\bibinfo {year} {2000})}\BibitemShut {NoStop}%
\bibitem [{\citenamefont {Bascompte}\ \emph {et~al.}(2005)\citenamefont
  {Bascompte}, \citenamefont {Meli{\'{a}}n},\ and\ \citenamefont
  {Sala}}]{Bascompte2005}%
  \BibitemOpen
  \bibfield  {author} {\bibinfo {author} {\bibfnamefont {J.}~\bibnamefont
  {Bascompte}}, \bibinfo {author} {\bibfnamefont {C.~J.}\ \bibnamefont
  {Meli{\'{a}}n}},\ and\ \bibinfo {author} {\bibfnamefont {E.}~\bibnamefont
  {Sala}},\ }\bibfield  {journal} {\bibinfo  {journal} {Proc. Natl. Acad. Sci.
  USA}\ }\href {https://doi.org/10.1073/pnas.0501562102}
  {10.1073/pnas.0501562102} (\bibinfo {year} {2005})\BibitemShut {NoStop}%
\bibitem [{\citenamefont {Ulanowicz}\ and\ \citenamefont
  {Baird}(1999)}]{Ulanowicz1999}%
  \BibitemOpen
  \bibfield  {author} {\bibinfo {author} {\bibfnamefont {R.~E.}\ \bibnamefont
  {Ulanowicz}}\ and\ \bibinfo {author} {\bibfnamefont {D.}~\bibnamefont
  {Baird}},\ }\bibfield  {journal} {\bibinfo  {journal} {Journal of Marine
  Systems}\ }\href {https://doi.org/10.1016/S0924-7963(98)90017-3}
  {10.1016/S0924-7963(98)90017-3} (\bibinfo {year} {1999})\BibitemShut
  {NoStop}%
\bibitem [{\citenamefont {Christian}\ and\ \citenamefont
  {Luczkovich}(1999)}]{Christian1999}%
  \BibitemOpen
  \bibfield  {author} {\bibinfo {author} {\bibfnamefont {R.~R.}\ \bibnamefont
  {Christian}}\ and\ \bibinfo {author} {\bibfnamefont {J.~J.}\ \bibnamefont
  {Luczkovich}},\ }\bibfield  {journal} {\bibinfo  {journal} {Ecological
  Modelling}\ }\href {https://doi.org/10.1016/S0304-3800(99)00022-8}
  {10.1016/S0304-3800(99)00022-8} (\bibinfo {year} {1999})\BibitemShut
  {NoStop}%
\bibitem [{\citenamefont {Goldwasser}\ and\ \citenamefont
  {Roughgarden}(1993)}]{Goldwasser1993}%
  \BibitemOpen
  \bibfield  {author} {\bibinfo {author} {\bibfnamefont {L.}~\bibnamefont
  {Goldwasser}}\ and\ \bibinfo {author} {\bibfnamefont {J.}~\bibnamefont
  {Roughgarden}},\ }\bibfield  {journal} {\bibinfo  {journal} {Ecology}\ }\href
  {https://doi.org/10.2307/1940492} {10.2307/1940492} (\bibinfo {year}
  {1993})\BibitemShut {NoStop}%
\bibitem [{\citenamefont {Huxham}\ \emph {et~al.}(1996)\citenamefont {Huxham},
  \citenamefont {Beaney},\ and\ \citenamefont {Raffaelli}}]{Huxham1996}%
  \BibitemOpen
  \bibfield  {author} {\bibinfo {author} {\bibfnamefont {M.}~\bibnamefont
  {Huxham}}, \bibinfo {author} {\bibfnamefont {S.}~\bibnamefont {Beaney}},\
  and\ \bibinfo {author} {\bibfnamefont {D.}~\bibnamefont {Raffaelli}},\
  }\bibfield  {journal} {\bibinfo  {journal} {Oikos}\ }\href
  {https://doi.org/10.2307/3546201} {10.2307/3546201} (\bibinfo {year}
  {1996})\BibitemShut {NoStop}%
\bibitem [{\citenamefont {Ekl{\"o}f}\ \emph {et~al.}(2013)\citenamefont
  {Ekl{\"o}f}, \citenamefont {Jacob}, \citenamefont {Kopp}, \citenamefont
  {Bosch}, \citenamefont {Castro-Urgal}, \citenamefont {Chacoff}, \citenamefont
  {Dalsgaard}, \citenamefont {de~Sassi}, \citenamefont {Galetti}, \citenamefont
  {Guimar{\~a}es} \emph {et~al.}}]{ref12}%
  \BibitemOpen
  \bibfield  {author} {\bibinfo {author} {\bibfnamefont {A.}~\bibnamefont
  {Ekl{\"o}f}}, \bibinfo {author} {\bibfnamefont {U.}~\bibnamefont {Jacob}},
  \bibinfo {author} {\bibfnamefont {J.}~\bibnamefont {Kopp}}, \bibinfo {author}
  {\bibfnamefont {J.}~\bibnamefont {Bosch}}, \bibinfo {author} {\bibfnamefont
  {R.}~\bibnamefont {Castro-Urgal}}, \bibinfo {author} {\bibfnamefont {N.~P.}\
  \bibnamefont {Chacoff}}, \bibinfo {author} {\bibfnamefont {B.}~\bibnamefont
  {Dalsgaard}}, \bibinfo {author} {\bibfnamefont {C.}~\bibnamefont {de~Sassi}},
  \bibinfo {author} {\bibfnamefont {M.}~\bibnamefont {Galetti}}, \bibinfo
  {author} {\bibfnamefont {P.~R.}\ \bibnamefont {Guimar{\~a}es}}, \emph
  {et~al.},\ }\href@noop {} {\bibfield  {journal} {\bibinfo  {journal} {Ecology
  letters}\ }\textbf {\bibinfo {volume} {16}},\ \bibinfo {pages} {577}
  (\bibinfo {year} {2013})}\BibitemShut {NoStop}%
\bibitem [{\citenamefont {Dunne}\ \emph {et~al.}(2008)\citenamefont {Dunne},
  \citenamefont {Williams}, \citenamefont {Martinez}, \citenamefont {Wood},\
  and\ \citenamefont {Erwin}}]{Dunne2008}%
  \BibitemOpen
  \bibfield  {author} {\bibinfo {author} {\bibfnamefont {J.~A.}\ \bibnamefont
  {Dunne}}, \bibinfo {author} {\bibfnamefont {R.~J.}\ \bibnamefont {Williams}},
  \bibinfo {author} {\bibfnamefont {N.~D.}\ \bibnamefont {Martinez}}, \bibinfo
  {author} {\bibfnamefont {R.~A.}\ \bibnamefont {Wood}},\ and\ \bibinfo
  {author} {\bibfnamefont {D.~H.}\ \bibnamefont {Erwin}},\ }\href
  {https://doi.org/10.1371/journal.pbio.0060102} {\bibfield  {journal}
  {\bibinfo  {journal} {PLoS Biology}\ }\textbf {\bibinfo {volume} {6}},\
  \bibinfo {pages} {693} (\bibinfo {year} {2008})}\BibitemShut {NoStop}%
\bibitem [{\citenamefont {Havens}(1992)}]{ref21}%
  \BibitemOpen
  \bibfield  {author} {\bibinfo {author} {\bibfnamefont {K.}~\bibnamefont
  {Havens}},\ }\href@noop {} {\bibfield  {journal} {\bibinfo  {journal}
  {Science}\ }\textbf {\bibinfo {volume} {257}},\ \bibinfo {pages} {1107}
  (\bibinfo {year} {1992})}\BibitemShut {NoStop}%
\bibitem [{\citenamefont {Martinez}(1991)}]{ref36}%
  \BibitemOpen
  \bibfield  {author} {\bibinfo {author} {\bibfnamefont {N.~D.}\ \bibnamefont
  {Martinez}},\ }\href@noop {} {\bibfield  {journal} {\bibinfo  {journal}
  {Ecological monographs}\ }\textbf {\bibinfo {volume} {61}},\ \bibinfo {pages}
  {367} (\bibinfo {year} {1991})}\BibitemShut {NoStop}%
\bibitem [{\citenamefont {Link}(2002)}]{ref73}%
  \BibitemOpen
  \bibfield  {author} {\bibinfo {author} {\bibfnamefont {J.}~\bibnamefont
  {Link}},\ }\href@noop {} {\bibfield  {journal} {\bibinfo  {journal} {Marine
  ecology progress series}\ }\textbf {\bibinfo {volume} {230}},\ \bibinfo
  {pages} {1} (\bibinfo {year} {2002})}\BibitemShut {NoStop}%
\bibitem [{\citenamefont {Warren}(1989)}]{ref74}%
  \BibitemOpen
  \bibfield  {author} {\bibinfo {author} {\bibfnamefont {P.~H.}\ \bibnamefont
  {Warren}},\ }\href@noop {} {\bibfield  {journal} {\bibinfo  {journal}
  {Oikos}\ ,\ \bibinfo {pages} {299}} (\bibinfo {year} {1989})}\BibitemShut
  {NoStop}%
\bibitem [{\citenamefont {Yodzis}(1998)}]{Yodzis1998}%
  \BibitemOpen
  \bibfield  {author} {\bibinfo {author} {\bibfnamefont {P.}~\bibnamefont
  {Yodzis}},\ }\bibfield  {journal} {\bibinfo  {journal} {Journal of Animal
  Ecology}\ }\href {https://doi.org/10.1046/j.1365-2656.1998.00224.x}
  {10.1046/j.1365-2656.1998.00224.x} (\bibinfo {year} {1998})\BibitemShut
  {NoStop}%
\bibitem [{\citenamefont {Ulanowicz}\ \emph {et~al.}(1998)\citenamefont
  {Ulanowicz}, \citenamefont {Bondavalli},\ and\ \citenamefont
  {Egnotovich}}]{ref32}%
  \BibitemOpen
  \bibfield  {author} {\bibinfo {author} {\bibfnamefont {R.~E.}\ \bibnamefont
  {Ulanowicz}}, \bibinfo {author} {\bibfnamefont {C.}~\bibnamefont
  {Bondavalli}},\ and\ \bibinfo {author} {\bibfnamefont {M.}~\bibnamefont
  {Egnotovich}},\ }\href@noop {} {\bibfield  {journal} {\bibinfo  {journal}
  {Annual Report to the United States Geological Service Biological Resources
  Division Ref. No.[UMCES] CBL}\ ,\ \bibinfo {pages} {98}} (\bibinfo {year}
  {1998})}\BibitemShut {NoStop}%
\bibitem [{\citenamefont {Cole}(1981)}]{ref44}%
  \BibitemOpen
  \bibfield  {author} {\bibinfo {author} {\bibfnamefont {B.~J.}\ \bibnamefont
  {Cole}},\ }\href@noop {} {\bibfield  {journal} {\bibinfo  {journal}
  {Science}\ }\textbf {\bibinfo {volume} {212}},\ \bibinfo {pages} {83}
  (\bibinfo {year} {1981})}\BibitemShut {NoStop}%
\bibitem [{\citenamefont {Lott}(1979)}]{ref45}%
  \BibitemOpen
  \bibfield  {author} {\bibinfo {author} {\bibfnamefont {D.~F.}\ \bibnamefont
  {Lott}},\ }\href@noop {} {\bibfield  {journal} {\bibinfo  {journal}
  {Zeitschrift f{\"u}r Tierpsychologie}\ }\textbf {\bibinfo {volume} {49}},\
  \bibinfo {pages} {418} (\bibinfo {year} {1979})}\BibitemShut {NoStop}%
\bibitem [{\citenamefont {Schein}\ and\ \citenamefont {Fohrman}(1955)}]{ref46}%
  \BibitemOpen
  \bibfield  {author} {\bibinfo {author} {\bibfnamefont {M.~W.}\ \bibnamefont
  {Schein}}\ and\ \bibinfo {author} {\bibfnamefont {M.~H.}\ \bibnamefont
  {Fohrman}},\ }\href@noop {} {\bibfield  {journal} {\bibinfo  {journal} {The
  British Journal of Animal Behaviour}\ }\textbf {\bibinfo {volume} {3}},\
  \bibinfo {pages} {45} (\bibinfo {year} {1955})}\BibitemShut {NoStop}%
\bibitem [{\citenamefont {Grant}(1973)}]{ref47}%
  \BibitemOpen
  \bibfield  {author} {\bibinfo {author} {\bibfnamefont {T.}~\bibnamefont
  {Grant}},\ }\href@noop {} {\bibfield  {journal} {\bibinfo  {journal} {Animal
  Behaviour}\ }\textbf {\bibinfo {volume} {21}},\ \bibinfo {pages} {449}
  (\bibinfo {year} {1973})}\BibitemShut {NoStop}%
\bibitem [{\citenamefont {Takahata}(1991)}]{ref48}%
  \BibitemOpen
  \bibfield  {author} {\bibinfo {author} {\bibfnamefont {Y.}~\bibnamefont
  {Takahata}},\ }\href@noop {} {\bibfield  {journal} {\bibinfo  {journal} {The
  monkeys of Arashiyama. State University of New York Press, Albany}\ ,\
  \bibinfo {pages} {123}} (\bibinfo {year} {1991})}\BibitemShut {NoStop}%
\bibitem [{\citenamefont {Clutton-Brock}\ \emph {et~al.}(1976)\citenamefont
  {Clutton-Brock}, \citenamefont {Greenwood},\ and\ \citenamefont
  {Powell}}]{ref49}%
  \BibitemOpen
  \bibfield  {author} {\bibinfo {author} {\bibfnamefont {T.}~\bibnamefont
  {Clutton-Brock}}, \bibinfo {author} {\bibfnamefont {P.}~\bibnamefont
  {Greenwood}},\ and\ \bibinfo {author} {\bibfnamefont {R.}~\bibnamefont
  {Powell}},\ }\href@noop {} {\bibfield  {journal} {\bibinfo  {journal}
  {Zeitschrift f{\"u}r Tierpsychologie}\ }\textbf {\bibinfo {volume} {41}},\
  \bibinfo {pages} {202} (\bibinfo {year} {1976})}\BibitemShut {NoStop}%
\bibitem [{\citenamefont {Hass}(1991)}]{ref50}%
  \BibitemOpen
  \bibfield  {author} {\bibinfo {author} {\bibfnamefont {C.~C.}\ \bibnamefont
  {Hass}},\ }\href@noop {} {\bibfield  {journal} {\bibinfo  {journal} {Journal
  of Zoology}\ }\textbf {\bibinfo {volume} {225}},\ \bibinfo {pages} {509}
  (\bibinfo {year} {1991})}\BibitemShut {NoStop}%
\bibitem [{\citenamefont {van Hooff}\ and\ \citenamefont
  {Wensing}(1987)}]{ref51}%
  \BibitemOpen
  \bibfield  {author} {\bibinfo {author} {\bibfnamefont {J.~A.}\ \bibnamefont
  {van Hooff}}\ and\ \bibinfo {author} {\bibfnamefont {J.~A.}\ \bibnamefont
  {Wensing}},\ }\href@noop {} {\bibfield  {journal} {\bibinfo  {journal}
  {Journal of Veterinary Behavior}\ } (\bibinfo {year} {1987})}\BibitemShut
  {NoStop}%
\bibitem [{\citenamefont {Garfield}()}]{ref17}%
  \BibitemOpen
  \bibfield  {author} {\bibinfo {author} {\bibfnamefont {E.}~\bibnamefont
  {Garfield}},\ }\href@noop {} {\bibinfo {title} {Index of citation networks
  produced by analyses from the software {HistCite}}},\ \bibinfo {howpublished}
  {\url{http://www.garfield.library.upenn.edu/histcomp/index.html}}\BibitemShut
  {NoStop}%
\bibitem [{\citenamefont {Hummon}\ and\ \citenamefont {Dereian}(1989)}]{ref16}%
  \BibitemOpen
  \bibfield  {author} {\bibinfo {author} {\bibfnamefont {N.~P.}\ \bibnamefont
  {Hummon}}\ and\ \bibinfo {author} {\bibfnamefont {P.}~\bibnamefont
  {Dereian}},\ }\href@noop {} {\bibfield  {journal} {\bibinfo  {journal}
  {Social networks}\ }\textbf {\bibinfo {volume} {11}},\ \bibinfo {pages} {39}
  (\bibinfo {year} {1989})}\BibitemShut {NoStop}%
\bibitem [{\citenamefont {Schubert}(2002)}]{ref14}%
  \BibitemOpen
  \bibfield  {author} {\bibinfo {author} {\bibfnamefont {A.}~\bibnamefont
  {Schubert}},\ }\href@noop {} {\bibfield  {journal} {\bibinfo  {journal}
  {Scientometrics}\ }\textbf {\bibinfo {volume} {53}},\ \bibinfo {pages} {3}
  (\bibinfo {year} {2002})}\BibitemShut {NoStop}%
\bibitem [{\citenamefont {Ley}(2002)}]{ref26}%
  \BibitemOpen
  \bibfield  {author} {\bibinfo {author} {\bibfnamefont {M.}~\bibnamefont
  {Ley}},\ }in\ \href@noop {} {\emph {\bibinfo {booktitle} {International
  symposium on string processing and information retrieval}}}\ (\bibinfo
  {organization} {Springer},\ \bibinfo {year} {2002})\ pp.\ \bibinfo {pages}
  {1--10}\BibitemShut {NoStop}%
\bibitem [{\citenamefont {Milo}\ \emph {et~al.}(2004)\citenamefont {Milo},
  \citenamefont {Itzkovitz}, \citenamefont {Kashtan}, \citenamefont {Levitt},
  \citenamefont {Shen-Orr}, \citenamefont {Ayzenshtat}, \citenamefont
  {Sheffer},\ and\ \citenamefont {Alon}}]{Milo2004}%
  \BibitemOpen
  \bibfield  {author} {\bibinfo {author} {\bibfnamefont {R.}~\bibnamefont
  {Milo}}, \bibinfo {author} {\bibfnamefont {S.}~\bibnamefont {Itzkovitz}},
  \bibinfo {author} {\bibfnamefont {N.}~\bibnamefont {Kashtan}}, \bibinfo
  {author} {\bibfnamefont {R.}~\bibnamefont {Levitt}}, \bibinfo {author}
  {\bibfnamefont {S.}~\bibnamefont {Shen-Orr}}, \bibinfo {author}
  {\bibfnamefont {I.}~\bibnamefont {Ayzenshtat}}, \bibinfo {author}
  {\bibfnamefont {M.}~\bibnamefont {Sheffer}},\ and\ \bibinfo {author}
  {\bibfnamefont {U.}~\bibnamefont {Alon}},\ }\bibfield  {journal} {\bibinfo
  {journal} {Science}\ }\href {https://doi.org/10.1126/science.1089167}
  {10.1126/science.1089167} (\bibinfo {year} {2004})\BibitemShut {NoStop}%
\bibitem [{\citenamefont {Krebs}()}]{ref42}%
  \BibitemOpen
  \bibfield  {author} {\bibinfo {author} {\bibfnamefont {V.}~\bibnamefont
  {Krebs}},\ }\href@noop {} {\bibinfo {title} {Madoff feeder funds.}},\
  \bibinfo {howpublished} {\url{
  http://www.thenetworkthinkers.com/2009/02/madoff-feeder-funds.html}}\BibitemShut
  {NoStop}%
\bibitem [{\citenamefont {Flandreau}\ and\ \citenamefont
  {Jobst}(2005)}]{ref31}%
  \BibitemOpen
  \bibfield  {author} {\bibinfo {author} {\bibfnamefont {M.}~\bibnamefont
  {Flandreau}}\ and\ \bibinfo {author} {\bibfnamefont {C.}~\bibnamefont
  {Jobst}},\ }\href@noop {} {\bibfield  {journal} {\bibinfo  {journal} {The
  Journal of Economic History}\ }\textbf {\bibinfo {volume} {65}},\ \bibinfo
  {pages} {977} (\bibinfo {year} {2005})}\BibitemShut {NoStop}%
\bibitem [{\citenamefont {De~Domenico}\ \emph {et~al.}(2015)\citenamefont
  {De~Domenico}, \citenamefont {Nicosia}, \citenamefont {Arenas},\ and\
  \citenamefont {Latora}}]{ref30}%
  \BibitemOpen
  \bibfield  {author} {\bibinfo {author} {\bibfnamefont {M.}~\bibnamefont
  {De~Domenico}}, \bibinfo {author} {\bibfnamefont {V.}~\bibnamefont
  {Nicosia}}, \bibinfo {author} {\bibfnamefont {A.}~\bibnamefont {Arenas}},\
  and\ \bibinfo {author} {\bibfnamefont {V.}~\bibnamefont {Latora}},\ }\href
  {https://doi.org/10.1038/ncomms7864} {\bibfield  {journal} {\bibinfo
  {journal} {Nature communications}\ }\textbf {\bibinfo {volume} {6}},\
  \bibinfo {pages} {6864} (\bibinfo {year} {2015})}\BibitemShut {NoStop}%
\bibitem [{\citenamefont {Smith}\ and\ \citenamefont {White}(1992)}]{ref60}%
  \BibitemOpen
  \bibfield  {author} {\bibinfo {author} {\bibfnamefont {D.~A.}\ \bibnamefont
  {Smith}}\ and\ \bibinfo {author} {\bibfnamefont {D.~R.}\ \bibnamefont
  {White}},\ }\href@noop {} {\bibfield  {journal} {\bibinfo  {journal} {Social
  Forces}\ }\textbf {\bibinfo {volume} {70}},\ \bibinfo {pages} {857} (\bibinfo
  {year} {1992})}\BibitemShut {NoStop}%
\bibitem [{\citenamefont {De~Nooy}\ \emph {et~al.}(2018)\citenamefont
  {De~Nooy}, \citenamefont {Mrvar},\ and\ \citenamefont {Batagelj}}]{ref61}%
  \BibitemOpen
  \bibfield  {author} {\bibinfo {author} {\bibfnamefont {W.}~\bibnamefont
  {De~Nooy}}, \bibinfo {author} {\bibfnamefont {A.}~\bibnamefont {Mrvar}},\
  and\ \bibinfo {author} {\bibfnamefont {V.}~\bibnamefont {Batagelj}},\
  }\href@noop {} {\emph {\bibinfo {title} {Exploratory social network analysis
  with Pajek: Revised and expanded edition for updated software}}},\
  Vol.~\bibinfo {volume} {46}\ (\bibinfo  {publisher} {Cambridge University
  Press},\ \bibinfo {year} {2018})\BibitemShut {NoStop}%
\bibitem [{\citenamefont {Asllani}\ and\ \citenamefont
  {Carletti}(2018)}]{Asllani2018PRE}%
  \BibitemOpen
  \bibfield  {author} {\bibinfo {author} {\bibfnamefont {M.}~\bibnamefont
  {Asllani}}\ and\ \bibinfo {author} {\bibfnamefont {T.}~\bibnamefont
  {Carletti}},\ }\href {https://doi.org/10.1103/PhysRevE.97.042302} {\bibfield
  {journal} {\bibinfo  {journal} {Phys. Rev. E}\ }\textbf {\bibinfo {volume}
  {97}},\ \bibinfo {pages} {042302} (\bibinfo {year} {2018})}\BibitemShut
  {NoStop}%
\bibitem [{\citenamefont {Muolo}\ \emph {et~al.}(2019)\citenamefont {Muolo},
  \citenamefont {Asllani}, \citenamefont {Fanelli}, \citenamefont {Maini},\
  and\ \citenamefont {Carletti}}]{Muolo2019}%
  \BibitemOpen
  \bibfield  {author} {\bibinfo {author} {\bibfnamefont {R.}~\bibnamefont
  {Muolo}}, \bibinfo {author} {\bibfnamefont {M.}~\bibnamefont {Asllani}},
  \bibinfo {author} {\bibfnamefont {D.}~\bibnamefont {Fanelli}}, \bibinfo
  {author} {\bibfnamefont {P.~K.}\ \bibnamefont {Maini}},\ and\ \bibinfo
  {author} {\bibfnamefont {T.}~\bibnamefont {Carletti}},\ }\href
  {https://doi.org/10.1016/j.jtbi.2019.07.004} {\bibfield  {journal} {\bibinfo
  {journal} {J. Theor. Biol.}\ }\textbf {\bibinfo {volume} {480}},\ \bibinfo
  {pages} {81} (\bibinfo {year} {2019})}\BibitemShut {NoStop}%
\bibitem [{\citenamefont {Nicoletti}\ \emph {et~al.}(2019)\citenamefont
  {Nicoletti}, \citenamefont {Fanelli}, \citenamefont {Zagli}, \citenamefont
  {Asllani}, \citenamefont {Battistelli}, \citenamefont {Carletti},
  \citenamefont {Chisci}, \citenamefont {Innocenti},\ and\ \citenamefont
  {Livi}}]{NN_stoch}%
  \BibitemOpen
  \bibfield  {author} {\bibinfo {author} {\bibfnamefont {S.}~\bibnamefont
  {Nicoletti}}, \bibinfo {author} {\bibfnamefont {D.}~\bibnamefont {Fanelli}},
  \bibinfo {author} {\bibfnamefont {N.}~\bibnamefont {Zagli}}, \bibinfo
  {author} {\bibfnamefont {M.}~\bibnamefont {Asllani}}, \bibinfo {author}
  {\bibfnamefont {G.}~\bibnamefont {Battistelli}}, \bibinfo {author}
  {\bibfnamefont {T.}~\bibnamefont {Carletti}}, \bibinfo {author}
  {\bibfnamefont {L.}~\bibnamefont {Chisci}}, \bibinfo {author} {\bibfnamefont
  {G.}~\bibnamefont {Innocenti}},\ and\ \bibinfo {author} {\bibfnamefont
  {R.}~\bibnamefont {Livi}},\ }\href {https://doi.org/10.1063/1.5099538}
  {\bibfield  {journal} {\bibinfo  {journal} {Chaos}\ }\textbf {\bibinfo
  {volume} {29}},\ \bibinfo {pages} {083123} (\bibinfo {year}
  {2019})}\BibitemShut {NoStop}%
\bibitem [{\citenamefont {Baggio}\ \emph {et~al.}(2020)\citenamefont {Baggio},
  \citenamefont {Rutten}, \citenamefont {Hennequin},\ and\ \citenamefont
  {Zampieri}}]{baggio2020efficient}%
  \BibitemOpen
  \bibfield  {author} {\bibinfo {author} {\bibfnamefont {G.}~\bibnamefont
  {Baggio}}, \bibinfo {author} {\bibfnamefont {V.}~\bibnamefont {Rutten}},
  \bibinfo {author} {\bibfnamefont {G.}~\bibnamefont {Hennequin}},\ and\
  \bibinfo {author} {\bibfnamefont {S.}~\bibnamefont {Zampieri}},\ }\href
  {https://doi.org/10.1126/sciadv.aba2282} {\bibfield  {journal} {\bibinfo
  {journal} {Sci. Adv.}\ }\textbf {\bibinfo {volume} {6}},\ \bibinfo {pages}
  {eaba2282} (\bibinfo {year} {2020})}\BibitemShut {NoStop}%
\bibitem [{\citenamefont {Johnson}(2020)}]{Johnson2020digraph}%
  \BibitemOpen
  \bibfield  {author} {\bibinfo {author} {\bibfnamefont {S.}~\bibnamefont
  {Johnson}},\ }\href {https://doi.org/10.1088/2632-072x/ab8e2f} {\bibfield
  {journal} {\bibinfo  {journal} {J. Phys. Complexity}\ }\textbf {\bibinfo
  {volume} {1}},\ \bibinfo {pages} {015003} (\bibinfo {year}
  {2020})}\BibitemShut {NoStop}%
\bibitem [{\citenamefont {Muolo}\ \emph {et~al.}(2021)\citenamefont {Muolo},
  \citenamefont {Carletti}, \citenamefont {Gleeson},\ and\ \citenamefont
  {Asllani}}]{Muolo2021}%
  \BibitemOpen
  \bibfield  {author} {\bibinfo {author} {\bibfnamefont {R.}~\bibnamefont
  {Muolo}}, \bibinfo {author} {\bibfnamefont {T.}~\bibnamefont {Carletti}},
  \bibinfo {author} {\bibfnamefont {J.~P.}\ \bibnamefont {Gleeson}},\ and\
  \bibinfo {author} {\bibfnamefont {M.}~\bibnamefont {Asllani}},\ }\bibfield
  {journal} {\bibinfo  {journal} {Entropy}\ }\textbf {\bibinfo {volume} {23}},\
  \href {https://doi.org/10.3390/e23010036} {10.3390/e23010036} (\bibinfo
  {year} {2021})\BibitemShut {NoStop}%
\bibitem [{\citenamefont {Duan}\ \emph {et~al.}(2022)\citenamefont {Duan},
  \citenamefont {Nishikawa}, \citenamefont {Eroglu},\ and\ \citenamefont
  {Motter}}]{Duan_Motter}%
  \BibitemOpen
  \bibfield  {author} {\bibinfo {author} {\bibfnamefont {C.}~\bibnamefont
  {Duan}}, \bibinfo {author} {\bibfnamefont {T.}~\bibnamefont {Nishikawa}},
  \bibinfo {author} {\bibfnamefont {D.}~\bibnamefont {Eroglu}},\ and\ \bibinfo
  {author} {\bibfnamefont {A.~E.}\ \bibnamefont {Motter}},\ }\href
  {https://doi.org/10.1126/sciadv.abm8310} {\bibfield  {journal} {\bibinfo
  {journal} {Science Advances}\ }\textbf {\bibinfo {volume} {8}},\ \bibinfo
  {pages} {eabm8310} (\bibinfo {year} {2022})},\ \Eprint
  {https://arxiv.org/abs/https://www.science.org/doi/pdf/10.1126/sciadv.abm8310}
  {https://www.science.org/doi/pdf/10.1126/sciadv.abm8310} \BibitemShut
  {NoStop}%
\bibitem [{\citenamefont {Pecora}\ and\ \citenamefont {Carroll}(1998)}]{MSF}%
  \BibitemOpen
  \bibfield  {author} {\bibinfo {author} {\bibfnamefont {L.~M.}\ \bibnamefont
  {Pecora}}\ and\ \bibinfo {author} {\bibfnamefont {T.~L.}\ \bibnamefont
  {Carroll}},\ }\href {https://doi.org/10.1103/PhysRevLett.80.2109} {\bibfield
  {journal} {\bibinfo  {journal} {Phys. Rev. Lett.}\ }\textbf {\bibinfo
  {volume} {80}},\ \bibinfo {pages} {2109} (\bibinfo {year}
  {1998})}\BibitemShut {NoStop}%
\bibitem [{\citenamefont {Zakharova}\ \emph {et~al.}(2014)\citenamefont
  {Zakharova}, \citenamefont {Kapeller},\ and\ \citenamefont
  {Sch\"oll}}]{amp_chimera}%
  \BibitemOpen
  \bibfield  {author} {\bibinfo {author} {\bibfnamefont {A.}~\bibnamefont
  {Zakharova}}, \bibinfo {author} {\bibfnamefont {M.}~\bibnamefont
  {Kapeller}},\ and\ \bibinfo {author} {\bibfnamefont {E.}~\bibnamefont
  {Sch\"oll}},\ }\href {https://doi.org/10.1103/PhysRevLett.112.154101}
  {\bibfield  {journal} {\bibinfo  {journal} {Phys. Rev. Lett.}\ }\textbf
  {\bibinfo {volume} {112}},\ \bibinfo {pages} {154101} (\bibinfo {year}
  {2014})}\BibitemShut {NoStop}%
\bibitem [{\citenamefont {Umbanhowar}\ \emph {et~al.}(1996)\citenamefont
  {Umbanhowar}, \citenamefont {Melo},\ and\ \citenamefont
  {Swinney}}]{umbanhowar_localized_1996}%
  \BibitemOpen
  \bibfield  {author} {\bibinfo {author} {\bibfnamefont {P.~B.}\ \bibnamefont
  {Umbanhowar}}, \bibinfo {author} {\bibfnamefont {F.}~\bibnamefont {Melo}},\
  and\ \bibinfo {author} {\bibfnamefont {H.~L.}\ \bibnamefont {Swinney}},\
  }\href {https://doi.org/10.1038/382793a0} {\bibfield  {journal} {\bibinfo
  {journal} {Nature}\ }\textbf {\bibinfo {volume} {382}},\ \bibinfo {pages}
  {793} (\bibinfo {year} {1996})}\BibitemShut {NoStop}%
\bibitem [{\citenamefont {Vanag}\ and\ \citenamefont
  {Epstein}(2004)}]{Vanag_Epstein}%
  \BibitemOpen
  \bibfield  {author} {\bibinfo {author} {\bibfnamefont {V.~K.}\ \bibnamefont
  {Vanag}}\ and\ \bibinfo {author} {\bibfnamefont {I.~R.}\ \bibnamefont
  {Epstein}},\ }\href {https://doi.org/10.1103/PhysRevLett.92.128301}
  {\bibfield  {journal} {\bibinfo  {journal} {Phys. Rev. Lett.}\ }\textbf
  {\bibinfo {volume} {92}},\ \bibinfo {pages} {128301} (\bibinfo {year}
  {2004})}\BibitemShut {NoStop}%
\bibitem [{\citenamefont {Schmidt}\ and\ \citenamefont
  {Avitabile}(2020)}]{oscillon_net}%
  \BibitemOpen
  \bibfield  {author} {\bibinfo {author} {\bibfnamefont {H.}~\bibnamefont
  {Schmidt}}\ and\ \bibinfo {author} {\bibfnamefont {D.}~\bibnamefont
  {Avitabile}},\ }\href {https://doi.org/10.1063/1.5135579} {\bibfield
  {journal} {\bibinfo  {journal} {Chaos: An Interdisciplinary Journal of
  Nonlinear Science}\ }\textbf {\bibinfo {volume} {30}},\ \bibinfo {pages}
  {033133} (\bibinfo {year} {2020})},\ \Eprint
  {https://arxiv.org/abs/https://doi.org/10.1063/1.5135579}
  {https://doi.org/10.1063/1.5135579} \BibitemShut {NoStop}%
\bibitem [{\citenamefont {Sun}\ \emph {et~al.}(2009)\citenamefont {Sun},
  \citenamefont {Bollt},\ and\ \citenamefont {Nishikawa}}]{Sun_2009}%
  \BibitemOpen
  \bibfield  {author} {\bibinfo {author} {\bibfnamefont {J.}~\bibnamefont
  {Sun}}, \bibinfo {author} {\bibfnamefont {E.~M.}\ \bibnamefont {Bollt}},\
  and\ \bibinfo {author} {\bibfnamefont {T.}~\bibnamefont {Nishikawa}},\ }\href
  {https://doi.org/10.1209/0295-5075/85/60011} {\bibfield  {journal} {\bibinfo
  {journal} {Europhysics Letters}\ }\textbf {\bibinfo {volume} {85}},\ \bibinfo
  {pages} {60011} (\bibinfo {year} {2009})}\BibitemShut {NoStop}%
\bibitem [{\citenamefont {Zhang}\ and\ \citenamefont
  {Motter}(2017)}]{Zhang_2018}%
  \BibitemOpen
  \bibfield  {author} {\bibinfo {author} {\bibfnamefont {Y.}~\bibnamefont
  {Zhang}}\ and\ \bibinfo {author} {\bibfnamefont {A.~E.}\ \bibnamefont
  {Motter}},\ }\href {https://doi.org/10.1088/1361-6544/aa8fe7} {\bibfield
  {journal} {\bibinfo  {journal} {Nonlinearity}\ }\textbf {\bibinfo {volume}
  {31}},\ \bibinfo {pages} {R1} (\bibinfo {year} {2017})}\BibitemShut {NoStop}%
\bibitem [{\citenamefont {Varshney}\ \emph {et~al.}(2011)\citenamefont
  {Varshney}, \citenamefont {Chen}, \citenamefont {Paniagua}, \citenamefont
  {Hall},\ and\ \citenamefont {Chklovskii}}]{varshney_structural_2011}%
  \BibitemOpen
  \bibfield  {author} {\bibinfo {author} {\bibfnamefont {L.~R.}\ \bibnamefont
  {Varshney}}, \bibinfo {author} {\bibfnamefont {B.~L.}\ \bibnamefont {Chen}},
  \bibinfo {author} {\bibfnamefont {E.}~\bibnamefont {Paniagua}}, \bibinfo
  {author} {\bibfnamefont {D.~H.}\ \bibnamefont {Hall}},\ and\ \bibinfo
  {author} {\bibfnamefont {D.~B.}\ \bibnamefont {Chklovskii}},\ }\href@noop {}
  {\bibfield  {journal} {\bibinfo  {journal} {PLoS Computational Biology}\
  }\textbf {\bibinfo {volume} {7}},\ \bibinfo {pages} {e1001066} (\bibinfo
  {year} {2011})}\BibitemShut {NoStop}%
\bibitem [{\citenamefont {Sorrentino}(2012)}]{sorrentino_synchronization_2012}%
  \BibitemOpen
  \bibfield  {author} {\bibinfo {author} {\bibfnamefont {F.}~\bibnamefont
  {Sorrentino}},\ }\href@noop {} {\bibfield  {journal} {\bibinfo  {journal}
  {New Journal of Physics}\ }\textbf {\bibinfo {volume} {14}},\ \bibinfo
  {pages} {033035} (\bibinfo {year} {2012})}\BibitemShut {NoStop}%
\bibitem [{Note1()}]{Note1}%
  \BibitemOpen
  \bibinfo {note} {However, the analysis can be extended to cases where the
  Laplacian matrix is not diagonalizable by utilizing Jordan blocks \cite
  {nondiagonal}.}\BibitemShut {Stop}%
\bibitem [{\citenamefont {Contemori}\ \emph {et~al.}(2016)\citenamefont
  {Contemori}, \citenamefont {Di~Patti}, \citenamefont {Fanelli},\ and\
  \citenamefont {Miele}}]{contemori_multiple-scale_2016}%
  \BibitemOpen
  \bibfield  {author} {\bibinfo {author} {\bibfnamefont {S.}~\bibnamefont
  {Contemori}}, \bibinfo {author} {\bibfnamefont {F.}~\bibnamefont {Di~Patti}},
  \bibinfo {author} {\bibfnamefont {D.}~\bibnamefont {Fanelli}},\ and\ \bibinfo
  {author} {\bibfnamefont {F.}~\bibnamefont {Miele}},\ }\href
  {https://doi.org/10.1103/PhysRevE.93.032317} {\bibfield  {journal} {\bibinfo
  {journal} {Physical Review E}\ }\textbf {\bibinfo {volume} {93}},\ \bibinfo
  {pages} {032317} (\bibinfo {year} {2016})}\BibitemShut {NoStop}%
\bibitem [{\citenamefont {Cross}\ and\ \citenamefont
  {Greenside}(2009)}]{cross_pattern_2009}%
  \BibitemOpen
  \bibfield  {author} {\bibinfo {author} {\bibfnamefont {M.}~\bibnamefont
  {Cross}}\ and\ \bibinfo {author} {\bibfnamefont {H.}~\bibnamefont
  {Greenside}},\ }\href@noop {} {\emph {\bibinfo {title} {Pattern formation and
  dynamics in nonequilibrium systems}}}\ (\bibinfo  {publisher} {Cambridge
  University Press},\ \bibinfo {address} {Cambridge, UK ; New York},\ \bibinfo
  {year} {2009})\ \bibinfo {note} {oCLC: ocn268793786}\BibitemShut {NoStop}%
\bibitem [{\citenamefont {Nakao}(2014)}]{nakao_complex_2014}%
  \BibitemOpen
  \bibfield  {author} {\bibinfo {author} {\bibfnamefont {H.}~\bibnamefont
  {Nakao}},\ }\href {https://doi.org/10.1140/epjst/e2014-02220-1} {\bibfield
  {journal} {\bibinfo  {journal} {The European Physical Journal Special
  Topics}\ }\textbf {\bibinfo {volume} {223}},\ \bibinfo {pages} {2411}
  (\bibinfo {year} {2014})}\BibitemShut {NoStop}%
\bibitem [{\citenamefont {Di~Patti}\ \emph {et~al.}(2018)\citenamefont
  {Di~Patti}, \citenamefont {Fanelli}, \citenamefont {Miele},\ and\
  \citenamefont {Carletti}}]{di_patti_ginzburg-landau_2018}%
  \BibitemOpen
  \bibfield  {author} {\bibinfo {author} {\bibfnamefont {F.}~\bibnamefont
  {Di~Patti}}, \bibinfo {author} {\bibfnamefont {D.}~\bibnamefont {Fanelli}},
  \bibinfo {author} {\bibfnamefont {F.}~\bibnamefont {Miele}},\ and\ \bibinfo
  {author} {\bibfnamefont {T.}~\bibnamefont {Carletti}},\ }\href
  {https://doi.org/10.1016/j.cnsns.2017.08.012} {\bibfield  {journal} {\bibinfo
   {journal} {Communications in Nonlinear Science and Numerical Simulation}\
  }\textbf {\bibinfo {volume} {56}},\ \bibinfo {pages} {447} (\bibinfo {year}
  {2018})}\BibitemShut {NoStop}%
\bibitem [{Note2()}]{Note2}%
  \BibitemOpen
  \bibinfo {note} {{Adding more rigor to this statement, for the case of a
  continuum medium, it can be shown and computed numerically that in the domain
  formed by the parameters and the wave vector, it exists a contiguous region,
  known as the ``stability balloon'' where the heterogeneous pattern is stable
  \cite {cross_pattern_2009}.}}\BibitemShut {Stop}%
\bibitem [{\citenamefont {Strayer}\ and\ \citenamefont
  {Cummins}(1980)}]{ref41}%
  \BibitemOpen
  \bibfield  {author} {\bibinfo {author} {\bibfnamefont {F.~F.}\ \bibnamefont
  {Strayer}}\ and\ \bibinfo {author} {\bibfnamefont {M.~S.}\ \bibnamefont
  {Cummins}},\ }\href@noop {} {\bibfield  {journal} {\bibinfo  {journal}
  {Dominance relations: an ethological view of human conflict and social
  interaction. Edinburgh: Livingstone}\ } (\bibinfo {year} {1980})}\BibitemShut
  {NoStop}%
\bibitem [{\citenamefont {Zhabotinsky}\ \emph {et~al.}(1995)\citenamefont
  {Zhabotinsky}, \citenamefont {Dolnik},\ and\ \citenamefont {Epstein}}]{zhab}%
  \BibitemOpen
  \bibfield  {author} {\bibinfo {author} {\bibfnamefont {A.~M.}\ \bibnamefont
  {Zhabotinsky}}, \bibinfo {author} {\bibfnamefont {M.}~\bibnamefont
  {Dolnik}},\ and\ \bibinfo {author} {\bibfnamefont {I.~R.}\ \bibnamefont
  {Epstein}},\ }\href {https://doi.org/10.1063/1.469932} {\bibfield  {journal}
  {\bibinfo  {journal} {The Journal of Chemical Physics}\ }\textbf {\bibinfo
  {volume} {103}},\ \bibinfo {pages} {10306} (\bibinfo {year} {1995})},\
  \Eprint {https://arxiv.org/abs/https://doi.org/10.1063/1.469932}
  {https://doi.org/10.1063/1.469932} \BibitemShut {NoStop}%
\bibitem [{\citenamefont {Asllani}\ \emph {et~al.}(2013)\citenamefont
  {Asllani}, \citenamefont {Biancalani}, \citenamefont {Fanelli},\ and\
  \citenamefont {McKane}}]{asllani_linear_2013}%
  \BibitemOpen
  \bibfield  {author} {\bibinfo {author} {\bibfnamefont {M.}~\bibnamefont
  {Asllani}}, \bibinfo {author} {\bibfnamefont {T.}~\bibnamefont {Biancalani}},
  \bibinfo {author} {\bibfnamefont {D.}~\bibnamefont {Fanelli}},\ and\ \bibinfo
  {author} {\bibfnamefont {A.~J.}\ \bibnamefont {McKane}},\ }\href
  {https://doi.org/10.1140/epjb/e2013-40570-8} {\bibfield  {journal} {\bibinfo
  {journal} {The European Physical Journal B}\ }\textbf {\bibinfo {volume}
  {86}},\ \bibinfo {pages} {476} (\bibinfo {year} {2013})}\BibitemShut
  {NoStop}%
\bibitem [{\citenamefont {Asllani}\ \emph {et~al.}(2014)\citenamefont
  {Asllani}, \citenamefont {Challenger}, \citenamefont {Pavone}, \citenamefont
  {Sacconi},\ and\ \citenamefont {Fanelli}}]{asllani_theory_2014}%
  \BibitemOpen
  \bibfield  {author} {\bibinfo {author} {\bibfnamefont {M.}~\bibnamefont
  {Asllani}}, \bibinfo {author} {\bibfnamefont {J.~D.}\ \bibnamefont
  {Challenger}}, \bibinfo {author} {\bibfnamefont {F.~S.}\ \bibnamefont
  {Pavone}}, \bibinfo {author} {\bibfnamefont {L.}~\bibnamefont {Sacconi}},\
  and\ \bibinfo {author} {\bibfnamefont {D.}~\bibnamefont {Fanelli}},\ }\href
  {https://doi.org/10.1038/ncomms5517} {\bibfield  {journal} {\bibinfo
  {journal} {Nature Communications}\ }\textbf {\bibinfo {volume} {5}},\
  \bibinfo {pages} {4517} (\bibinfo {year} {2014})}\BibitemShut {NoStop}%
\bibitem [{\citenamefont {Murray}(2008)}]{Murray2008}%
  \BibitemOpen
  \bibfield  {author} {\bibinfo {author} {\bibfnamefont {J.~D.}\ \bibnamefont
  {Murray}},\ }\href {https://doi.org/10.1007/b98869} {\emph {\bibinfo {title}
  {Mathematical Biology II - Spatial Models and Biomedical Applications}}}\
  (\bibinfo  {publisher} {Springer-Verlag},\ \bibinfo {year}
  {2008})\BibitemShut {NoStop}%
\bibitem [{Note3()}]{Note3}%
  \BibitemOpen
  \bibinfo {note} {Hereby, by ``strongly stable'' we mean that the stability
  indicators as the Jacobian eigenvalue with the largest real part,
  respectively, the Maximum Lyapunov Exponent have a considerably large
  magnitude apart from being negative.}\BibitemShut {Stop}%
\bibitem [{\citenamefont {Strogatz}\ and\ \citenamefont
  {Dichter}(2018)}]{strogatz_book}%
  \BibitemOpen
  \bibfield  {author} {\bibinfo {author} {\bibfnamefont {S.}~\bibnamefont
  {Strogatz}}\ and\ \bibinfo {author} {\bibfnamefont {M.}~\bibnamefont
  {Dichter}},\ }\href@noop {} {\emph {\bibinfo {title} {Nonlinear {Dynamics}
  and {Chaos} with {Student} {Solutions} {Manual}: {With} {Applications} to
  {Physics}, {Biology}, {Chemistry}, and {Engineering}}}},\ \bibinfo {edition}
  {second edition (combined)}\ ed.\ (\bibinfo  {publisher} {CRC Press},\
  \bibinfo {address} {Boca Raton, FL},\ \bibinfo {year} {2018})\ \bibinfo
  {note} {oCLC: 1105704612}\BibitemShut {NoStop}%
\bibitem [{\citenamefont {Wright}\ \emph {et~al.}(2019)\citenamefont {Wright},
  \citenamefont {Yoon}, \citenamefont {Ferreira}, \citenamefont {Mendes},\ and\
  \citenamefont {Goltsev}}]{wright_central_2019}%
  \BibitemOpen
  \bibfield  {author} {\bibinfo {author} {\bibfnamefont {E.~A.~P.}\
  \bibnamefont {Wright}}, \bibinfo {author} {\bibfnamefont {S.}~\bibnamefont
  {Yoon}}, \bibinfo {author} {\bibfnamefont {A.~L.}\ \bibnamefont {Ferreira}},
  \bibinfo {author} {\bibfnamefont {J.~F.~F.}\ \bibnamefont {Mendes}},\ and\
  \bibinfo {author} {\bibfnamefont {A.~V.}\ \bibnamefont {Goltsev}},\ }\href
  {https://doi.org/10.1038/s41598-019-49537-8} {\bibfield  {journal} {\bibinfo
  {journal} {Scientific Reports}\ }\textbf {\bibinfo {volume} {9}},\ \bibinfo
  {pages} {13162} (\bibinfo {year} {2019})}\BibitemShut {NoStop}%
\bibitem [{Note4()}]{Note4}%
  \BibitemOpen
  \bibinfo {note} {{In principle, a SCC connected with the rest of the network
  with only outgoing links, can in principle synchronize as shown in \cite
  {symm_break} and consequently decrease the disordered oscillations of the
  leaders.}}\BibitemShut {Stop}%
\bibitem [{\citenamefont {Nishikawa}\ and\ \citenamefont
  {Motter}(2006)}]{nondiagonal}%
  \BibitemOpen
  \bibfield  {author} {\bibinfo {author} {\bibfnamefont {T.}~\bibnamefont
  {Nishikawa}}\ and\ \bibinfo {author} {\bibfnamefont {A.~E.}\ \bibnamefont
  {Motter}},\ }\href {https://doi.org/10.1103/PhysRevE.73.065106} {\bibfield
  {journal} {\bibinfo  {journal} {Phys. Rev. E}\ }\textbf {\bibinfo {volume}
  {73}},\ \bibinfo {pages} {065106} (\bibinfo {year} {2006})}\BibitemShut
  {NoStop}%
\end{thebibliography}%
\bibliographystyle{apsrev4-2}

\clearpage

%%%%%%%%%%%%%%%%%%%%%%%%%%%%%%%%%%%%%%%%%%%%%%%%%%%%%%%%%%%%%%%%%%%%%%%%%%%%%%%%%%%%%%%%%%%%%%%%%%%%%%%%%%%%%%%%%%%%%%%%%%%%%%%%%%%%%%%%%%%%%%%%%%%%%%%%%%%%%%%%%%%%%%%%%%%%%%%%%%%%%%%%%%%%%%%%

\section*{Supplementary Material}

\subsection{Effects of non-normality in the chimera pattern formation mechanism}
\label{sec:AppC}
\noindent

\begin{figure*}%[h!]
	\centering
	\includegraphics[width = \linewidth]{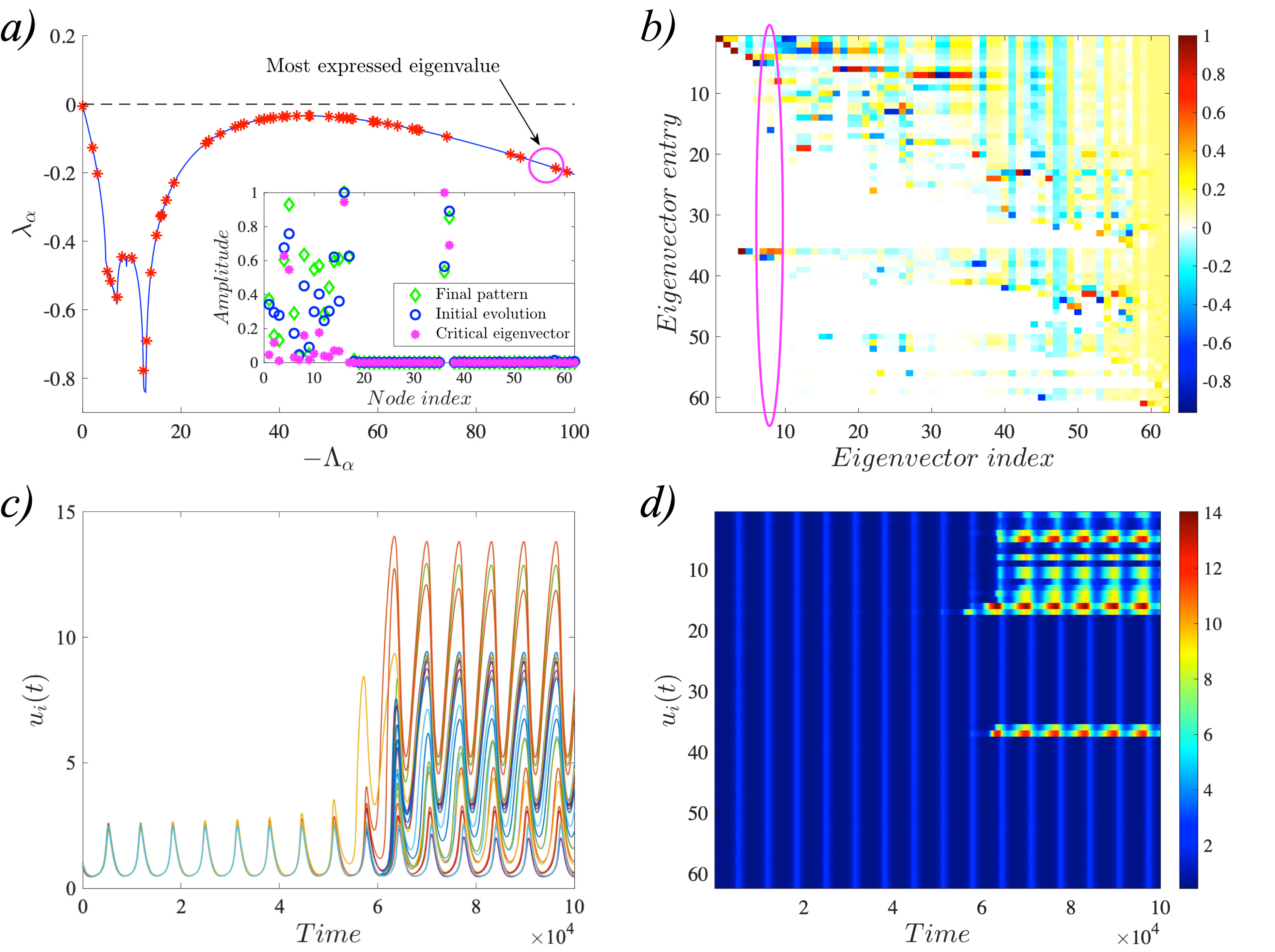}
	\caption{\textbf{Non-normality driven chimera patterns.} $\textbf{a)}$ The Master Stability Function (red stars) for the \textit{dominance among macaques} network \cite{ref31}. The most expressed eigenvalue correspond to the eigth eigenvector which is clearly farer from the threshold of instability than other modes. $\textbf{b)}$ Comparison between the most expressed eigenvector (magenta stars) and the normalized amplitudes of the initial evolution (blue circles) and the final pattern (green diamonds). Notice that although quantitatively very similar the comparison is less neat than previously. $\textbf{c)}$ The time series for each nodes where due to many nodes involved it is almost impossible to establish the shape of the pattern obtained. $\textbf{d)}$ The colormap representation of the evolution of the amplitude chimera state where the colorbar quantifies the magnitude of the patterns. The parameters for the Brusselator model are $b=2.5$, $c=1$, $D_u=0.007$, and $D_v=0.083$.}
	\label{Fig4}
\end{figure*}

The results shown {in the main text} are based on a symmetry breaking mechanism which allows for the selection of the shape of the final pattern by exploiting the spectral properties of the non-normal networks. Nevertheless, throughout this work so far, we have paid particular attention (except in Sec.~\ref{sec:III}) in choosing sufficiently small perturbations in order to allow the linear stability analysis to work appropriately in the initial regime of the pattern evolution. As emphasized in the main text this is crucial in the scenario of non-normal systems where uniform synchronized or fixed point states tend to have a very small basin of attraction making techniques such as the Master Stability Function difficult to implement if not handled with particular care. In fact, as shown in numerous previous studies \cite{Asllani2018PRE, Muolo2019, Muolo2021}, the linear approach ``might fail'' when the perturbation is not sufficiently small causing a non-normality driven instability in contrast to the analytical prediction, based on a linear stability study, suggesting that the system should indeed be stable. To understand the shape of the final patterns in this later scenario, we now analyze the \emph{dominance among macaques} network consisting of $62$ nodes where the non-normality is stronger due to the larger size of the network \cite{asllani2018structure}. Considering again the Brusselator model, we notice that although the MSF is strictly stable, Fig.~\ref{Fig4} $a)$, in panel $c)$ and in particular in panel $d)$ we can observe the emergence of amplitude chimera states. Not having a definite unstable mode makes the prediction of the final pattern apparently impossible. Nevertheless, when compared to the set of eigenvectors of the Laplacian matrix (see the inset of panel $a)$) there is a good comparison between the final pattern and its initial evolution with the eighth eigenvector. A plausible explanation for that from the physical point of view is based on the transient growth which prevails in the linear regime of stable non-normal system and drives the evolution of the perturbation as described by the formula $\delta x_i^{\gamma}=\sum_{\alpha=1}^{\Omega} \xi_{\alpha}^{\gamma}(t)\Phi_i^{(\alpha)}$. According to such a mechanism the mass will first accumulate at the nodes with higher magnitude of the fastest eigenvector(s) entries making the norm of the variables associated to the nodes $||\mathbf{x}||$ transiently larger independently of the globally dissipative system \cite{Asllani2018PRE}. 
An increasing of such a norm will also increase the nonlinear terms, thus stabilizing the system at some other, this time non-uniform, stable state. On the other side understanding what are the ``fastest'' eigenvectors involved in the linear component of the dynamics, is a non trivial task at all. In fact, from the global perspective, we know that despite the small perturbations, the system has been initialized outside the basin of the fully synchronized state. Although the tangent hyperplane spanned by the Laplacian eigenvectors is still a good approximation of the phase space at the local level, the shape of the basin of attraction (which might be unevenly distributed) (see Fig.  5 in Ref. \cite{Muolo2021}), the choice of the initial values, and the direction of the eigenvectors are expected to together play a role in the selection of the shape of the final pattern. For instance, the reason that the eighth eigenvector is more expressed might be partially due to the fact that it is more localized compared to the other ones, as it can be seen in Fig. \ref{Fig4} $b)$. However, not knowing a priori the shape of the basin of attraction and how this is related to the initial values, i.e., the functions $\xi_{\alpha}^{\gamma}(t)$, makes the prediction impossible based on the local analysis. Hence one should expect that the pattern formation in this case to be a trade-off between these different ingredients. Despite the lack of predictive power, the main point we want to make here is that an almost triangular shape of the matrix of the Laplacian eigenvectors will still yield an amplitude chimera pattern as a result, as can be clearly observed in Fig. \ref{Fig4}.  
%\onecolumngrid

%\twocolumngrid

\subsection{Supplementary Data}
\label{sec:AppB}
In the following Fig. \ref{FigSupp}, we {show how the triangularity in the adjacency and Laplacian matrices manifest in many empirical networks taken from different domains. For comparision base, we have also associated to each of the case below their normalized Henrici's departure from the non-normality index \cite{Trefethen2005}. All the data shown in the following have not gone any preprocess work apart from the relabelling of the nodes to emphasize their triangularity through the same algorithm explained in the main text.}

\begin{figure*}
	\centering
	\includegraphics[width = 0.85\linewidth]{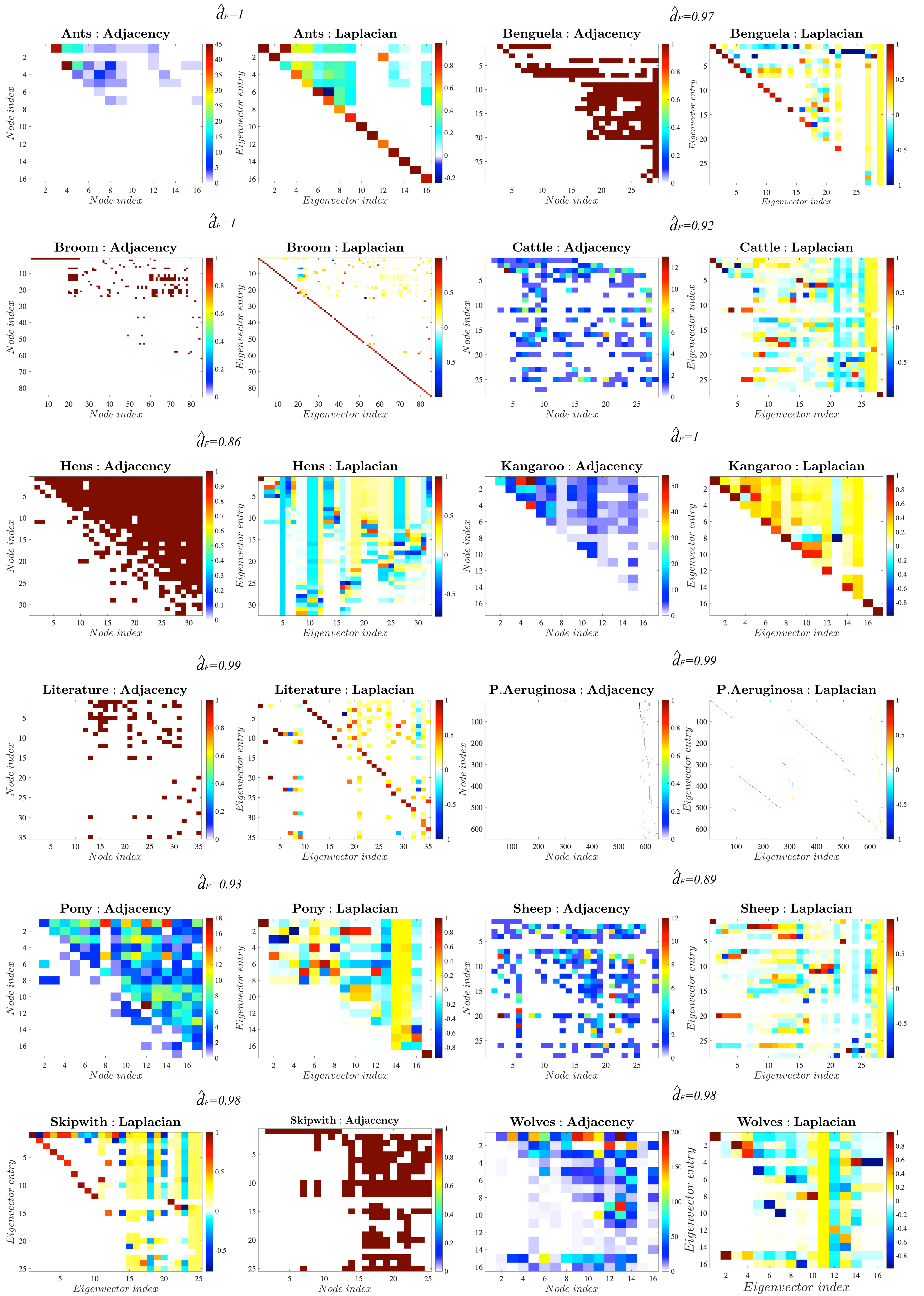}
	\caption{\textbf{Adjacency matrices and Laplacian eigenvectors of selected empirical networks.} We present the colormaps of the adjacency matriced and Laplacian eigenvectors of several empirical network which represents food webs, hierarchical organization between animals and human social networks (as referred in the titles of the panels). As shown, most of the networks manifest a strong triangular shape further strengthening our result. All the networks presented here have been hierarchally relabelled as explained in the main text.}
	\label{FigSupp}
\end{figure*}

\end{document}